\documentclass[acmsmall]{acmart}

\AtBeginDocument{
}

\setcopyright{cc}
\setcctype{by}
\acmJournal{PACMHCI}
\acmYear{2026}
\acmVolume{10}
\acmNumber{6}
\acmArticle{CSCW153}
\acmMonth{10}
\acmDOI{10.1145/3817001}

\usepackage{enumitem}

\usepackage{booktabs}
\usepackage{multirow}
\usepackage{graphicx}
\usepackage{listings}
\begin{document}

\title[Situated Practice Systems]{Situated Practice Systems: A Computational System for Supporting the Coaching and Practice of Regulation Skills for Innovation Work}

\author{Kapil Garg}
\orcid{0000-0003-4593-4766}
\affiliation{
\institution{University of California, Irvine}
\city{Irvine}
\state{California}
\country{USA}
\postcode{92697}
}
\affiliation{
\institution{Northwestern University}
\streetaddress{633 Clark St}
\city{Evanston}
\state{Illinois}
\country{USA}
\postcode{60208}
}
\email{kapilg@uci.edu}

\author{Darren Gergle}
\orcid{0000-0003-4052-0214}
\affiliation{
\institution{Northwestern University}
\streetaddress{633 Clark St}
\city{Evanston}
\state{Illinois}
\country{USA}
\postcode{60208}
}
\email{dgergle@northwestern.edu}

\author{Haoqi Zhang}
\orcid{0000-0002-7640-0532}
\affiliation{
\institution{Northwestern University}
\streetaddress{633 Clark St}
\city{Evanston}
\state{Illinois}
\country{USA}
\postcode{60208}
}
\email{hq@northwestern.edu}

\renewcommand{\shortauthors}{Kapil Garg, Darren Gergle, \& Haoqi Zhang}

\begin{abstract}
Students are increasingly expected to prepare for open-ended innovation work, which requires well-developed cognitive, metacognitive, and emotional regulation skills. College learning environments offer opportunities to work on real-world problems---such as in design and engineering---but students often remain unaware of their ineffective work practices and recurring regulation challenges, and may struggle to improve. Coaching from experts can help, but students' practices and regulation behaviors are largely invisible from work artifacts alone and are difficult to diagnose and track without computational support. We introduce {\em Situated Practice Systems (SPS)}, which provide: (1) an {\em Interactive Context-Assessment-Plan (CAP)} Notes tool to support coaches' understanding and modeling students' regulation-informed practices, and (2) {\em Practice Agents} that help students develop more effective practices. SPS uses {\em Practice Objects} to represent practices and regulation behaviors computationally, and {\em Practice Scripts} to automatically present suggested practices to students in relevant situations. In a formative 3-week field study, SPS helped coaches identify recurring regulation gaps and provide tailored practices. SPS also guided students in adopting more effective ways of working on their own and with others. We demonstrate how CSCW systems and learning environments can be designed to support the development of students' work practices and regulation skills, enabling them to lead innovation work.
\end{abstract}



\begin{teaserfigure}
  \centering
  \includegraphics[width=\textwidth]{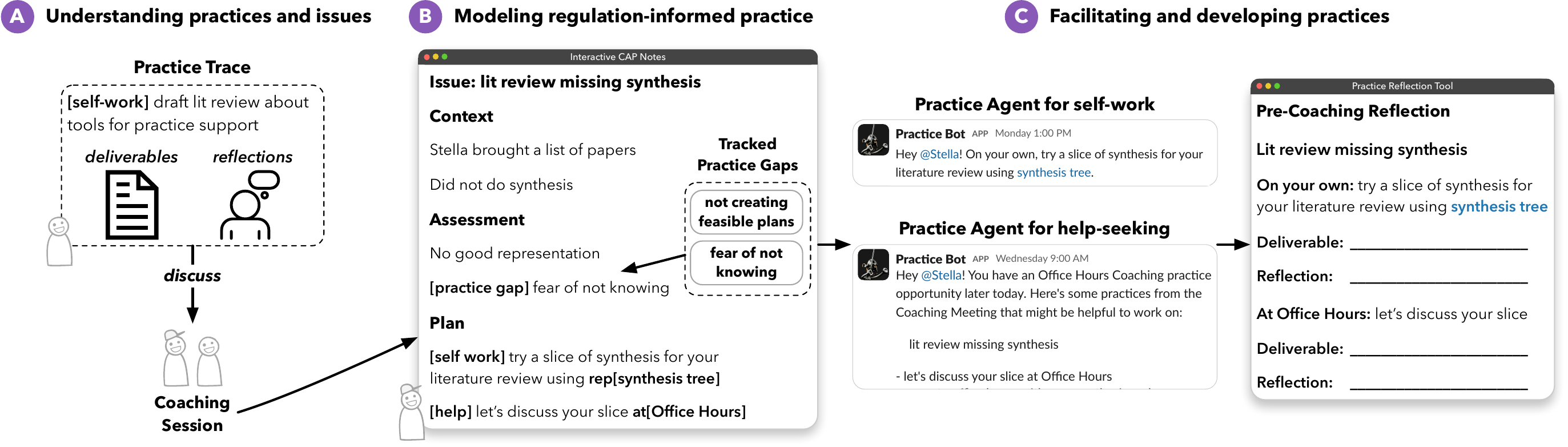}
  \caption{A Situated Practice System (SPS) computationally supports a coach in understanding, modeling, and facilitating students' innovation work practices. During coaching sessions, it provides the coach with Practice Traces of a student's prior practice (A). The coach can then model the student's practice based on discussions with the student and by drawing connections between specific work issues and the student's recurring Tracked Practice Gaps (B). Following coaching sessions, SPS provides Practice Agents that help students enact practices on their own and with others, and that prompt them to reflect afterward (C).}
  \label{fig:cap-teaser}
  \Description[Situated Practice System workflow for understanding, modeling, and facilitating practice.]{Three panels from left to right. Panel A, 'Understanding practices and issues', shows a Practice Trace for a student's self-work activity to draft a literature review, with an icon for a deliverable and reflection. This feeds into a coaching session where it is discussed with the coach. Panel B, 'Modeling regulation-informed practice', shows the Interactive Context-Assessment-Plan Notes tool interface, which displays the issue a coach is describing; Context; Assessment, which includes a box for Tracked Practice Gaps that can be added to an assessment note; and Plan. The issue is 'lit review missing synthesis'; the Context is that Stella, the student, brought a list of papers but did not do synthesis; the Assessment is that she did not have a good representation, with a Tracked Practice Gap of 'fear of not knowing'; and Plan with a self-work and help-seeking at office hours practice. Panel C, 'Facilitating and developing practices', shows Practice Agents sending self-work and help-seeking messages with Slack, followed by a pre-coaching reflection tool interface with separate deliverable and reflection fields for each practice.}
\end{teaserfigure}

\received{May 13, 2025}
\received[revised]{January 13, 2026}
\received[accepted]{March 17, 2026}

\maketitle

\section{Introduction}

There is an urgent need to train college students to tackle complex, open-ended {\em innovation work} in design, research, STEM, and entrepreneurship~\cite{jonassen_instructional_1997}. Innovation problems have vaguely structured goals and constraints, and often multiple solutions, if any exist~\cite{jonassen_instructional_1997,simon_structure_1973}. Students must learn to gradually uncover details about the problem space, continually adapting their work strategies as their work situations change. To effectively self-direct, students must develop {\em regulation skills}~\cite{zimmerman_investigating_2008}---cognitive, metacognitive, emotional, and strategic behaviors---that allow them to assess their current knowledge, choose where to focus efforts, determine effective strategies, carry out plans on their own or with help from others~\cite{jarvela_new_2013}, and reflect on their progress~\cite{zimmerman_becoming_2002}. 

Despite this need, college classrooms rarely provide the practice of {\em learning to innovate}, which is crucial for preparing students to design and implement innovative solutions to social and technical problems upon graduation~\cite{gerber2012extracurricular}. Project-based learning (PBL) presents a potential solution, as it can provide students with authentic practice on real-world problems~\cite{dunlap_problem-based_2005,dym_engineering_2005,frank2004integrating,prince2006inductive}---such as in design~\cite{puntambekar_toward_2005, rees_lewis_opportunities_2019}, engineering~\cite{dunlap_problem-based_2005}, and research~\cite{zhang2017agile}. However, PBL environments often overemphasize project progress. Feedback focuses on troubleshooting project-related issues, while {\em regulation gaps}---patterns of ineffective ways of working with oneself and others---can remain hidden and unaddressed. For instance, students may receive feedback on how to refine their prototype, but not on how they ignore key project risks during team meetings and tend to overwork rather than appropriately scope the work. Such ineffective ways of working often recur; troubleshooting surface-level issues overlooks the underlying regulation gaps shared across them, resulting in students failing to practice new ways of working that apply in these and other work situations. In other words, even when we provide students with real-world problems, we risk students not building the practices and regulation skills needed to lead innovation work on their own.

While experienced coaches can help students develop effective practices and regulation skills, doing so is challenging and time-consuming~\cite{huang_intelligent_2023}. Coaches already face challenges when providing project-level feedback on students' work outputs (e.g., prototypes in a design class). To teach regulation skills, coaches need to understand how students practice and regulate to produce these outputs, which is largely opaque from work artifacts alone. For instance, a prototype may be missing a key feature because the student did not realize it was needed (cognitive), overfocused on less important features (metacognitive), or reacted to fears that got in the way of their working on it (emotional). Each of these regulation challenges warrants different strategies that coaches must tailor to the student and their specific work needs, such as suggesting planning strategies that prioritize project risks or helping the student understand and work with their fears. 

This coaching task is challenging even at a moderate scale~\cite{rees_lewis_opportunities_2019}. Coaches need to understand each student's regulation gaps across work issues and track them over time to make connections between work issues and recurring regulation gaps in increasingly large classrooms. Moreover, even if coaches provide practices that address practice and regulation issues, students may struggle to enact them if they misunderstand how to apply a regulation skill, lack sufficient scaffolding, or have other unaddressed regulation gaps that make it difficult to enact the practice~\cite{rees_lewis_encouraging_2023,garg_networked-orchestration_2022,huang_intelligent_2023}.

Despite these challenges, there are few computational tools for supporting coaches and students to develop effective innovation practices and relevant regulation skills. While many coaches and students take notes during meetings, their plain-text notes remain ``static'', meaning they cannot be followed up on or tracked by computers. Existing tools in the classroom~\cite{lewis_planning_2018, sterman_kaleidoscope_2023} and the workplace~\cite{trello, asana, jira} largely track tasks and their outcomes, as opposed to the practices or regulation skills that a student needs to develop. Moreover, workflow tools for orchestrating classroom~\cite{dillenbourg_flexibility_2007, fischer_scripting_2007, kobbe_specifying_2007, weinberger_computer-supported_2009, miller_scripting_2015} or workplace activities~\cite{winograd_languageaction_1987, medina-mora_action_1992, van2013business, georgakopoulos_overview_1995, garg_orchestraton-scripts_2023} are pre-scripted and thus cannot be easily tailored to each student's specific work situations and practice needs. Without computational support, even experienced coaches often overlook critical, underdeveloped skills that students possess, and students struggle to develop their practice, slowing progress and learning. 

\begin{figure}
  \centering
  \includegraphics[width=\textwidth]{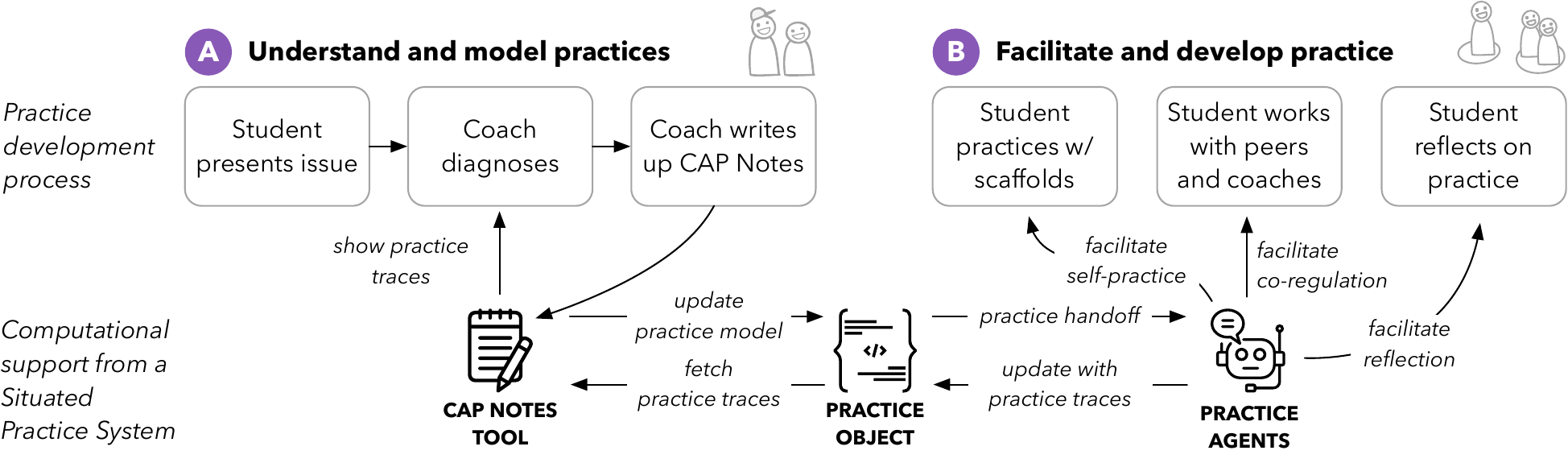}
  \caption{A Situated Practice System supports students in developing their innovation work practices in two ways. First (A), the Interactive CAP Notes tool supports coaches in understanding and modeling students' practices by showing students' prior Practice Traces and Tracked Practice Gaps, and scaffolding structured note-taking. Second (B), Practice Agents support students practicing on their own or with others, and encourage reflection after practice. The Practice Object acts as an intermediary, evolving representation of the student's practice that is handed off to Practice Agents to facilitate practice and updated with Practice Traces for later review.}
  \label{fig:cap-service-diagram}
  \Description[Service diagram for coaching practice and computational support from a Situated Practice System.]{A two-part workflow with the practice-development process on top and computational support below. Panel A, 'Understand and model practices', shows a student presenting an issue, a coach diagnosing it, and the coach writing CAP Notes. Panel B, 'Facilitate and develop practice', shows a student practicing with scaffolds, working with peers and coaches, and reflecting on practice. The lower row shows CAP Notes, the Practice Object, and Practice Agents connected by arrows. The CAP Notes tool shows Practice Traces fetched from the Practice Object to the coach during diagnoses, and updates the Practice Object when the coach takes notes. Then it updates the Practice Object, which passes practice information to the Practice Agent. The Practice Agent facilitates self-practice, co-regulation, and reflection, updating the Practice Object with Practice Traces afterward.}
\end{figure}

In this paper, we introduce {\em Situated Practice Systems (SPS)}, a system designed to help students develop effective work practices with the support of their coaches; see Figure~\ref{fig:cap-service-diagram}. An SPS supports a coach in understanding and modeling practices and regulation gaps with a structured notetaking tool called {\em Interactive Context-Assessment-Plan (CAP) Notes}. CAP Notes show coaches the outcomes and student reflections from previously suggested practices, helping them understand work issues and practice gaps---including recurring gaps that SPS tracks for each student---and suggest new regulation-informed practices to the student. Then, SPS supports students in enacting these practices through {\em Practice Agents} that automatically surface practice opportunities, such as practicing on their own with relevant scaffolds (e.g., design argument templates; see Appendix~\ref{appendix:template}) and with peers or mentors (e.g., during a peer help session or at Office Hours). Following these practices, Practice Agents prompt students to reflect on their practice, and the regulation skills they employed.

Our core conceptual contribution is {\em using computation to support understanding, modeling, and facilitating regulation skill practice to advance innovation work}. Historically, CSCW has used computation to provide support at the {\em task-layer} by tracking tasks, work outputs, and the people or resources that support them~\cite{ackerman_answer_1990, ackerman_answer_1996, mcdonald_expertise_2000, hui_introassist_2018}. Without a way to represent the practices underlying work tasks, we cannot build software to support the {\em practice-layer} that helps enact effective practices and track their development. SPS provides a computational representation of the student's practice and regulation gaps {\em through which} the coach can capture and track the student's development and recruit software agents to facilitate aspects of the practice outside of coaching sessions. 

We make two technical contributions through SPS. To make practices computationally representable and trackable, we created {\em Practice Objects}: a computational representation of a student's evolving practice that captures patterns of recurring behaviors across work situations. Across coaching sessions, a student's Practice Object expands to include new work issues, regulation gaps, and tailored practice suggestions, rather than only their tasks and work outputs. To enable Practice Agents to act on a coach's suggested practices, we introduce a technique for interpreting and compiling them into {\em Practice Scripts} that automatically facilitate both the practice and subsequent reflection. Furthermore, these scripts generate {\em Practice Traces} that capture practice outcomes and student reflection and store them in the Practice Object for later review by coaches.

Through a 3-week formative study in a research learning community, we demonstrate that CAP Notes supported coaches in understanding ineffective practices that led to work-related issues, identifying their underlying regulation gaps, and modeling regulation-informed practices for students to attempt. Practice Agents supported students in implementing suggested practices by providing resources and guidance from peers and mentors between coaching sessions, and helped students to reflect on and understand how new practices addressed their regulation challenges. These findings demonstrate how Situated Practice Systems explicitly support the adoption of more effective ways of working and the development of regulation skills needed for innovation work.

\section{Background}
Our work aims to provide computational support for coaching and enacting {\em regulation-informed practices} in innovation work. We review the literature on innovation work and the core regulation skills that contribute to students' success, the challenges of developing these skills, and the limitations of technology in supporting their development.

\subsection{Regulation Skills and Innovation Work}
Learning to innovate requires students to become capable of self-directing complex work~\cite{carlson_defining_2018,shah_agile_2023,lewis_planning_2018,rees_lewis_encouraging_2023,lewis2024research,lewis2020logic,zhang2017agile, garg_networked-orchestration_2022}. Ill-structured innovation problems~\cite{jonassen_instructional_1997,buchanan_wicked_1992} require students to {\em regulate} their learning and work process by assessing the current state of their knowledge, choosing where to focus effort, determining strategies for making progress, carrying out plans, reflecting on progress, and effectively engaging in help-seeking and collaboration as needed~\cite{jarvela_new_2013,zimmerman_becoming_2002}.\footnote{We use the term ``regulation'' from self-regulated learning (SRL) in the learning sciences (e.g., ~\cite{zimmerman1989models, zimmerman_becoming_2002}), where a student monitors, controls, and adapts their learning process toward a goal. While they share terminology and underlying mechanisms, this differs from self-regulation in psychology, which refers to the person's self-control over their thoughts, emotions, attention, or impulses to align their behaviors with a standard~\cite{baumeister1996selfregulation, vohs_handbook_2016}. We focus on understanding how students develop a capacity to become independent practitioners on open-ended problems (akin to a self-directed learner), which requires regulating their own learning and work processes and co-regulating with others in a learning or work community~\cite{jarvela_new_2013}.} Students engaging in authentic innovation work must develop a wide range of {\em regulation skills}, including: 

\begin{itemize}
  \item {\em Cognitive skills} for approaching problems with an unknown answer, or even knowing what the problem is. This includes: representing problem and solution spaces~\cite{shah_agile_2023,carlson2020design}; assessing risks~\cite{carlson_defining_2018}; critical thinking and argumentation; and core domain-specific methods for advancing problems and solutions (e.g., in design: prototyping, and user testing).
  \item {\em Metacognitive skills and dispositions}~\cite{hacker1998metacognition, mccormick2003metacognition} in the areas of planning, help-seeking, collaboration, and reflection~\cite{ambrose_how_2010, dunning2012self, hacker2000test}. This includes: forming feasible plans~\cite{zhang2017agile,chi_self-explanations_1989, schoenfeld1987s}; planning effective iterations~\cite{lewis_planning_2018,rees_lewis_encouraging_2023,lewis2024research}; leveraging resources and seeking help~\cite{jonassen_toward_2000, nelson1981help, pintrich_conceptual_2004, ryan_avoiding_2001}; communicating and working with teammates, peers, and coaches~\cite{jarvela_new_2013}; and awareness of one's own skills, abilities, and metacognitive blockers~\cite{schon1987educating, brown_psychological_1996,ambrose_how_2010, dunning2012self, hacker2000test}. 
  \item {\em Emotional regulation and dispositions toward self and learning} that affect one's motivation, cognition, and metacognition. These include understanding one's fears and anxieties and how one deals with failure~\cite{jarvenoja2018capturing,li_exploring_2024,mayer2001emotional,caruso2004emotional}, embracing challenges and learning from them~\cite{edmondson2012teaming,dweck2006mindset}, and embracing one's own independence when leading work~\cite{ambrose_how_2010,lepper2013motivational}.
\end{itemize}

College students who effectively regulate their learning are more likely to succeed in innovation work by: exploring problem and solution spaces effectively~\cite{carlson2020design,shah_agile_2023}; engaging in design iterations that quickly advance understanding~\cite{rees_lewis_encouraging_2023,lewis2024research}; and better planning for and responding to challenges and unknowns~\cite{lewis_planning_2018,zhang2017agile}. However, students are often unaware of their own work practices and the value of regulating learning across cognitive, metacognitive, and emotional dimensions. Moreover, students cannot easily develop effective practice on their own~\cite{collins_cognitive_1991,jonassen_all_2008,metcalfe_role_1994,flavell_metacognition_1979}. For example, consider a student in a design capstone who overworks themselves each week to create incomplete high-fidelity (hi-fi) software prototypes rather than low-fidelity (lo-fi) ones to test new designs. This student lacks strategies for {\em slicing} work to address project risks,\footnote{Throughout this paper, ``slicing'' refers to implementing a minimal, complete chunk of work that provides end-to-end understanding. The idea comes from Agile software methodologies that prioritize flexible, incremental development versus rigid, waterfall approaches~\cite{martin2003agile, cockburn_agile_2002}. In design, this may involve focusing on a user need, developing a low-fidelity prototype, {\em and} testing it on a few users; in research, it may entail selecting a scoped hypothesis, designing a formative experiment, {\em and} analyzing the results. Our studied research community (see Section~\ref{sec:dbr-method-setting}) places a strong emphasis on slicing, as it is an important strategy for planning effective iterations~\cite{lewis_planning_2018,rees_lewis_encouraging_2023,lewis2024research} that quickly advance understanding.} prioritizing the development of software over new understanding about the problem or solution space. To become effective innovators, students need experienced coaches to guide product and process development, but, most critically, to diagnose and address underlying regulation gaps that underpin their work practices. 

\subsection{Challenges for Developing Regulation Skills}
\label{ssec:reg-skill-challenges}

Coaches can support students in developing regulation skills~\cite{collins_cognitive_1989,hmelo-silver_goals_2006,hmelo-silver_problem-based_2004}, but doing so requires uncovering students' {\em regulation gaps}, which is often challenging and time-intensive~\cite{huang_intelligent_2023}. Consider the same design student who brings in an incomplete prototype to a coaching session. A {\em product-focused} coach might help the student identify a missing key feature and keep them on track toward building a complete prototype. A {\em process-focused} coach may investigate why the student struggled to complete the prototype, helping the student recognize that they did not need to build out fully functioning software at this time. This can help the student learn a better approach to their current work situation (e.g., lo-fi prototyping) that the product-focused coach overlooked. 

A {\em regulation-focused} coach delves even deeper, uncovering underlying gaps in metacognition and emotional regulation that impede the student from including a key feature they know is important. For example, this coach may find that: (1) the student associates delivering value with producing functioning software; and (2) the student tends to lose sight of what is actually important when overwhelmed, and defaults to simply cranking out work (building software). In other words, the underlying issue is not that the student did not build a lo-fi prototype, but rather that (1) the student lacked a framework for planning design iterations that focuses on delivering learning over product~\cite{lewis_planning_2018,rees_lewis_encouraging_2023,lewis2024research}, and that (2) their ``just-get-it-done'' mentality when overwhelmed impedes their considering and adopting more effective strategies for reaching their learning and work goals. Understanding and addressing such regulation gaps can help the student to resolve their current work issue, {\em and} to develop new strategies, mental models, and attitudes (i.e., regulation skills) that help them to innovate more effectively and efficiently {\em across} work situations~\cite{van2017ten,argyris1997organizational}.

While uncovering regulation gaps is crucial to helping students become more effective innovators, it remains challenging and time-consuming, even for experienced coaches. Prior study of university engineering design coaches~\cite{rees_lewis_opportunities_2019,rees_lewis_encouraging_2023}, research mentors~\cite{garg_networked-orchestration_2022}, and entrepreneurship coaches~\cite{huang_intelligent_2023} revealed that coaches struggle to: (1) elicit gaps in students' practice from work artifacts and conversation alone; (2) build a model of the student's practice to identify and address those gaps; and (3) track students' practice across coaching sessions. Our needfinding extends these findings, revealing challenges in how coaches uncover practice gaps from work products and conversations, link work issues to regulation gaps, and facilitate practices. These difficulties significantly limit coaching effectiveness, often exacerbated by the number of students and teams coaches manage~\cite{rees_lewis_opportunities_2019}. 

Moreover, students often struggle to enact the practices their coaches suggest~\cite{rees_lewis_encouraging_2023,garg_networked-orchestration_2022,huang_intelligent_2023}. Students may misunderstand how to apply a regulation strategy during practice or forget it altogether, resulting in deliverables that differ greatly from their coaches' expectations. In our needfinding, we observed that students often needed structured scaffolds (e.g., use a design argument template; see Appendix~\ref{appendix:template}) and additional support from peers or coaches. While coaches often suggest such scaffolds and help opportunities, students can overlook or fail to apply them in ways that address their specific regulation gaps. Consequently, while students may make some progress on their projects, they often fail to address key project risks stemming from unresolved regulation gaps.

\subsection{Our Approach: Situated Practice Systems for Regulation Skill Development}

In summary, advancing students' capability for innovation work requires advancing the coaching of regulation skills, which remains difficult without technological support. Despite advances in multimodal learning analytics~\cite{jarvela2024triggers} and AI tools for coaching ill-structured problems~\cite{fournier-viger_its-ill-def_2010,lamnina2018invention}, we currently cannot automate the process of understanding students' innovation work practices, diagnosing regulation gaps, and providing tailored practice opportunities. At the same time, intelligent mentoring tools to support such coaching remain scarce. Coaches develop models of how each student practices and identify regulation gaps, but may not document these in a structured way that allows them to integrate computational affordances for tracking and facilitating practices outside of coaching sessions, or helping identify recurring gaps across work issues.

Our work introduces {\em Situated Practice Systems} that provide computational support for developing regulation skills. Existing tools in CSCW largely provide support at the {\em task-layer}, such as for task tracking~\cite{trello, asana, jira} and expertise finding~\cite{ackerman_answer_1990, ackerman_answer_1996, mcdonald_expertise_2000, hui_introassist_2018}. Instead, Situated Practice Systems provide support at the {\em practice-layer} through an {\em Interactive CAP Notes} interface for coaches to understand students' regulation behaviors within and across coaching sessions and to model new practices to resolve those gaps, and {\em Practice Agents} for students to facilitate these practices.

There are two technical challenges to realizing Situated Practice Systems. First, existing tools offer no computational means to represent work practices and regulation behaviors so they can be tracked, facilitated, and reflected upon by coaches and students across work situations, coaching sessions, and practice activities. Existing tools for project-based learning track students' work plans~\cite{lewis_planning_2018}, artifacts, and reflections~\cite{sterman_kaleidoscope_2023}, but not the details of how students practice and how that may impact their work outputs. Similarly, workplace tools from CSCW support task tracking~\cite{trello, asana, jira}, agenda building~\cite{raikundalia_web_1998, schummer_towards_2009, vivacqua_agendabuilder_2010}, synthesizing in-meeting conversations~\cite{rahman_mixtape_2020, chiu_liteminutes_2001, foster_cognoter_1986}, and following up after meetings~\cite{wang_meeting_2024}. However, these systems focus on task management and conversations around tasks (e.g., what workers did, what workers need to do, and what to prioritize), but not on regulation skills necessary to complete tasks (e.g., the representation used to structure thinking and challenges in regulating emotions amid failure) or how disparate tasks a person is struggling with may share the same underlying regulation gap. We address this by creating a {\em Practice Object} abstraction that ties specific regulation behaviors to specific work issues that arise, which helps coaches capture and reveal {\em patterns} in a student's practice and regulation. These, in turn, inform the formation of {\em regulation-informed practices} that coaches suggest to address both the current work issue and the underlying regulation gaps that recur across work situations. Each student's Practice Object evolves over weeks, providing a longitudinal view of their innovation work practice.

Second, existing tools do not computationally encode {\em practice activities}, such as working with peers. Practices are typically documented in plain-text notes, limiting software agents' ability to facilitate them. While collaboration scripts for learning~\cite{dillenbourg_flexibility_2007, fischer_scripting_2007, kobbe_specifying_2007, weinberger_computer-supported_2009, miller_scripting_2015} and workflow builders for productivity~\cite{winograd_languageaction_1987, medina-mora_action_1992, van2013business, georgakopoulos_overview_1995, garg_orchestraton-scripts_2023} can encode and facilitate activities, they are designed for recurring, general activities (e.g., generally reflecting; sharing deliverables). While regulation strategies are shared across work situations (e.g., planning strategies are relevant for lo-fi prototyping and user testing), coaches must tailor practices to students' specific work situations and regulation gaps, making it cumbersome and sometimes intractable to write custom scripts for each student across different projects. To resolve this challenge, we introduce techniques to generate on-demand {\em Practice Scripts} that convert text-based representations of activities a coach wants a student to attempt into programs that run at opportune times to promote those practices and prompt reflections afterward.

\section{Method and Setting}
\label{sec:dbr-method-setting}
We followed a design-based research process~\cite{easterday2018logic} over six months to develop our design arguments for how computational tools can support the development of regulation skills for innovation work. The first month focused on observational needfinding in coaching sessions (Section~\ref{sec:needfinding}). Months 2--5 introduced prototypes as design probes~\cite{gaver1999design}, with weekly iterations. We describe the final prototype and design iterations in Section~\ref{sec:cap-system}. In the final month, we conducted a 3-week formative study to understand how our prototype addressed the obstacles identified in needfinding (Sections~\ref{sec:formative-study-methods}--\ref{sec:formative-study}).

We situated our study in the Design, Technology, and Research (DTR) program at Northwestern University,\footnote{https://dtr.northwestern.edu/} a community of practice for undergraduate and graduate research training that implements the Agile Research Studios (ARS) model~\cite{zhang2017agile}. We chose this setting because self-directing research work is highly complex and requires students to develop sophisticated work strategies and regulation skills for dealing with uncertainty and open-endedness that are broadly applicable for innovation work~\cite{margaryan_self-regulated_2013}. Moreover, such skill training is a key focus of the community (see the DTR Annual Letters~\cite{zhang2022dtr, zhang2023dtr, zhang2024dtr, zhang2025dtr}), which allows us to understand how coaching is done and how technology may help. Finally, ARS provides a rich learning ecosystem that resembles the networked, collaborative workplaces many students will join after graduation~\cite{garg_networked-orchestration_2022}. During our study, two faculty members coached 13--16 undergraduate, master's, and PhD students, organized into 7--9 project teams of at most 2 students each. Faculty held 5 1-hour coaching sessions per week, with each session accommodating up to 2 teams. Two of the paper's authors were a faculty member and a PhD student in DTR; the PhD student did not participate in the study.

\section{Needfinding: How Expert Coaches Coach Regulation-Informed Practice}
\label{sec:needfinding}

\begin{figure}[t]
  \centering
  \includegraphics[width=\textwidth]{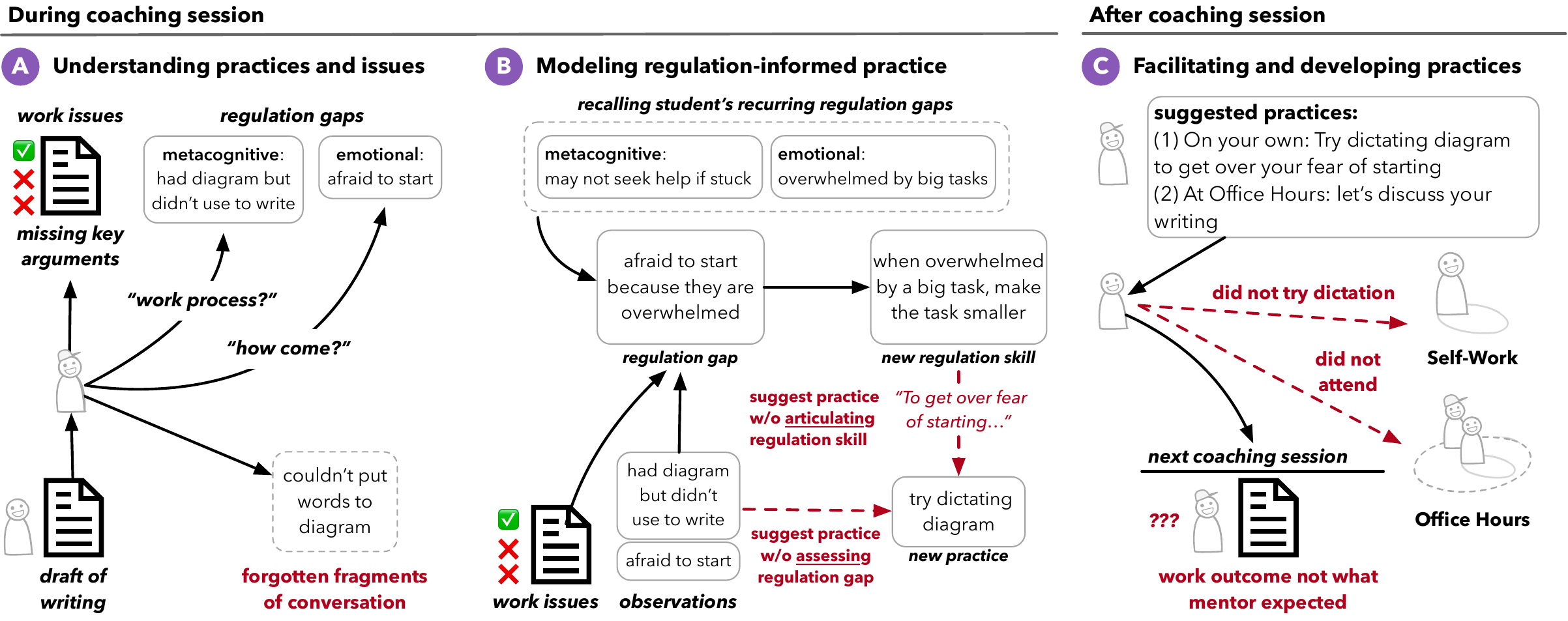}
  \caption{When coaching a practice, coaches can struggle to: (A) understand (gaps in) a student's practice from work artifacts and conversation alone, since eliciting practices and regulation gaps require extensive prompting from the coach and fragments of conversation useful for diagnosis can be forgotten; (B) model the links between their observations, recurring regulation gaps in a student's practices, the needed skills, and a new practice; and (C) facilitate a student's practice across self-work and co-regulated interactions with peers and coaches, resulting in students not practicing as expected and generating unexpected work outcomes.}
  \label{fig:needfinding}
  \Description[Three coaching challenges across understanding, modeling, and facilitating practice.]{Three panels are split during coaching (Panel A and Panel B) and after coaching (Panel C). Panel A, 'Understanding practices and issues', begins with a student presenting a work artifact and shows the coach asking about the work process and why the issue occurred. The inquiry surfaces gaps in metacognitive and emotional regulation, but also leaves forgotten fragments of conversation. Panel B, 'Modeling regulation-informed practice', shows a clockwise flow of four parts. From the bottom left, work issues and observations identify a regulation gap, which informs a new regulation skill, which in turn informs a new practice. The regulation gap is informed by prior, recurring gaps the coach recalls. Warning arrows mark failure cases in which a coach suggests practice without assessing the regulation gap or articulating the new regulation skill. Panel C, 'Facilitating and developing practices', shows the suggested practices for the student to do on their own and during office hours. The student neither attempts the self-work practice nor comes to office hours, so the next coaching session begins with a work outcome the mentor did not expect.}
\end{figure}

While prior work has studied coaches' challenges in coaching regulation skills and students' struggles to practice them (see Section~\ref{ssec:reg-skill-challenges}), we sought to better understand how coaching conversations unfold so that tools can support them. The lead author observed 1--3 coaching sessions each week, selecting different sessions to cover various projects and students. During each observation, the researcher noted how coaches discussed research progress and deliverables with students; the questions asked to identify work issues and regulation gaps; how new practices were modeled; and outcomes of suggested practices. The researcher also met weekly with one coach to discuss the observations and solicit feedback on prototypes. As shown in Figure~\ref{fig:needfinding}, we identified three challenges for coaches: (A) understanding practices and potential issues; (B) modeling a regulation-informed practice; and (C) facilitating and developing students' practices after coaching sessions.

\subsection{Understanding Practices and Issues}

Coaches could readily identify issues in the work artifacts students produced, but struggled to uncover the ineffective practices that led to those issues; see Figure~\ref{fig:needfinding}a. For example, a coach reviewing a student's draft quickly noticed it was missing a key argument they had previously discussed. To uncover what went wrong, the coach asked the student about their work process (e.g., did they remember to include the argument? Did they have a structure to describe it?) and any obstacles (e.g., if they did, why did they not try to write it?). While this discussion eventually revealed a regulation gap, it consumed over 20 of the 30 minutes available during the coaching session, leaving little time to model and coach a new practice~\cite{huang_intelligent_2023, rees_lewis_opportunities_2019}. 

Additionally, coaches often found {\em signals} of potential practice issues before pinpointing the actual issue amid conversation. These were often fragments of ideas that might be issues, such as a strategy the student avoided or a regulation skill they may not have applied. While they offer important clues about the true practice gaps, they are not easy to track, as coaches also ask additional questions and discuss project or regulation-related topics with the student. 

To help coaches understand how prior practices unfolded, Situated Practice Systems: (1) \textbf{capture {\em Practice Traces} that connect the work artifact to students' practices that produced it}; and (2) \textbf{provide loose structures for tracking signals of ineffective work practices that may arise in conversation}, which a coach can follow up on as practice issues become clearer~\cite{piolat_cognitive_2005}.

\subsection{Modeling Regulation-Informed Practices}
Coaches also faced challenges in modeling regulation-informed practices for students to try; see Figure~\ref{fig:needfinding}b. Continuing with the earlier example, the coach noticed that the student had a good diagram of what to write, but did not convert it into words. To understand why, the coach drew connections between the current work issue and the student's recurring regulation gaps, such as their fears of starting something unknown. Making these non-trivial connections led the coach to recognize that the student was afraid to start writing because they felt overwhelmed. The coach then suggested an alternative practice: dictating their (already effective) diagram and transcribing the recording to help overcome the initial fear of writing something new. However, at other times, coaches bypassed diagnosing the regulation gap and suggested practices to address the work issue directly. Even when coaches considered potential regulation gaps, they sometimes failed to incorporate the regulation skill into their suggested practices. Despite best intentions, both cases resulted in coaches giving students practices that were not regulation-informed, where the core regulation skill to apply in that specific work situation and the reasoning for why it would help resolve their regulation gap were not communicated, affecting how well the students could practice.

To help coaches model regulation-informed practices, Situated Practice Systems provide \textbf{structured representations for modeling the links among a work issue, (recurring) regulation gaps in the practice, and proposed practices} that address regulation gaps in the context of the specific work issue. These structures help externalize a coach's diagnosis and practice model, not just the observed work issues, and develop over time as students' practices evolve or recur.

\subsection{Facilitating and Developing Practices}

Even when coaches proposed practices, students often struggled to enact them; Figure~\ref{fig:needfinding}c. In our observations, coaches suggested practices for various work and regulation needs, but students struggled to understand how to practice when opportunities arose. For example, coaches would suggest scaffolds that students did not use while working, such as a prior diagram structure for writing the key arguments about a project. In other cases, students forgot a suggested practice or needed more scaffolding to do it, such as help from an unfamiliar peer. Moreover, coaches lacked the bandwidth to facilitate such practices in the moment, such as before help-seeking opportunities.

To help facilitate practices outside of coaching sessions, Situated Practice Systems \textbf{provide in-situ support for enacting a practice} through lightweight computational representations that coaches can use to express practices during coaching sessions. With a computational representation of the practice, the system can surface relevant practices to students at opportune moments, such as when they re-plan after a coaching session or on the morning of a peer help session. It can also include helpful practice scaffolds, such as structured representations to guide thinking. After these practices, the system can prompt students to reflect on the regulation skills they applied and then share them with coaches, along with any work artifacts, before the next coaching session.

\section{Situated Practice Systems to Support Regulation-Informed Practice Development}
\label{sec:cap-system}
We developed {\em Situated Practice Systems} to computationally support understanding, modeling, and facilitating regulation-informed practices. We first describe the interactional model of the system---\textbf{Interactive Context-Assessment-Plan (CAP) Notes} for the coach and \textbf{Practice Agents} for students---and then its technical implementation through Practice Objects and Practice Scripts.

\subsection{Interactive CAP Notes: A Structured Notetaking Tool for Coaches}

\subsubsection{Understanding Practices and Issues}

\begin{figure}[t]
  \centering
  \includegraphics[width=\textwidth]{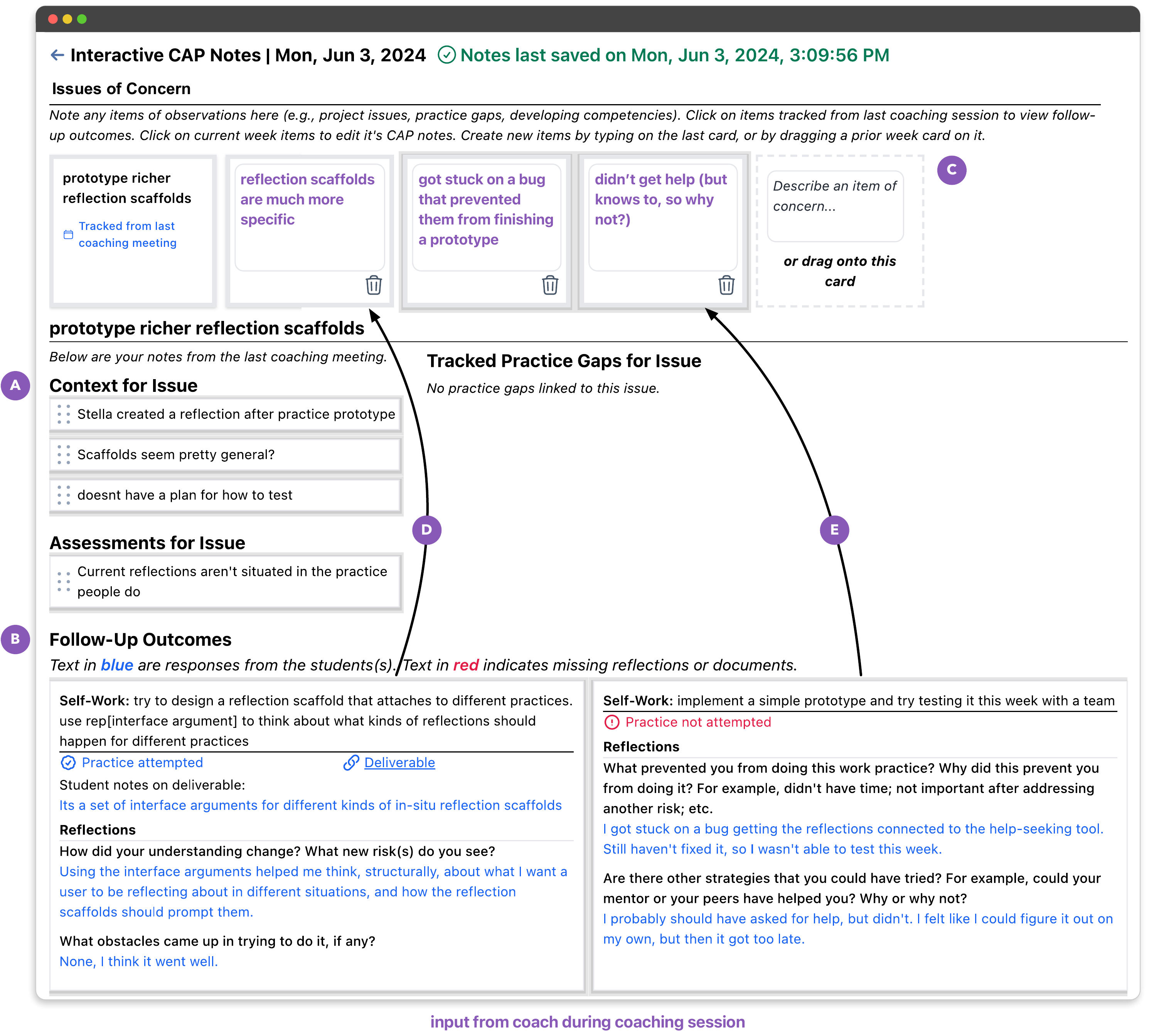}
  \caption{CAP Notes provides coaches with a Practice Trace of the prior practice suggestion (A) with the student's deliverables and practice reflections included (B). During review and discussion with the student, coaches can capture their observations as Issues of Concern (C), such as effective practices (D) and potential issues (E).}
  \label{fig:cap-notes-eliciting}
  \Description[Homepage user interface of the Interactive Context-Assessment-Plan (CAP) Notes tool for a project team.]{The interface has two parts. The top area contains a row of Issues of Concern cards side by side. One card showing the prior week's issue is selected; three other cards have been edited following review of the Practice Trace from the selected issue. The area below is the Practice Trace for the selected issue. The top half of this section is divided into two columns: the left column includes Context and Assessments, and the right column includes Tracked Practice Gaps. Below are the Follow-Up Outcomes, with two cards placed side by side. The left card shows a self-work practice the student completed, while the right card shows one they did not complete.}
\end{figure}

When a coach opens CAP Notes, they see a {\em Practice Trace} with their suggested practices and the student's practice outcome; see Figure~\ref{fig:cap-notes-eliciting}. Before a coaching session, students use a reflection tool to share their deliverables and reflect on their practice. Practice Traces show coaches their prior Context and Assessments to aid recall (A), with the students' deliverables and reflections underneath (B), including if practices were not done (as self-reported) or if deliverables or reflections were missing. If relevant, CAP Notes pulls live data from workplace tools, such as a student's planning log if the coach suggested a re-planning practice. These traces help the coach begin to see the practices that led to students' current work outcomes.

While reviewing Practice Traces and later during the coaching session, CAP Notes provides loose structures called {\em Issues of Concern} to help coaches capture observations (C). Here, they can note effective practices or work outcomes (D) and signals of issues to discuss (i.e., project needs, ways of working, and regulation skills) without requiring them to explicate what they are yet (E). Loose structures help coaches track their discussions with students about work practices and provide enough structure to refine their understanding of practice issues.

\subsubsection{Modeling New Regulation-Informed Practices}

\begin{figure}[t]
  \centering
  \includegraphics[width=\textwidth]{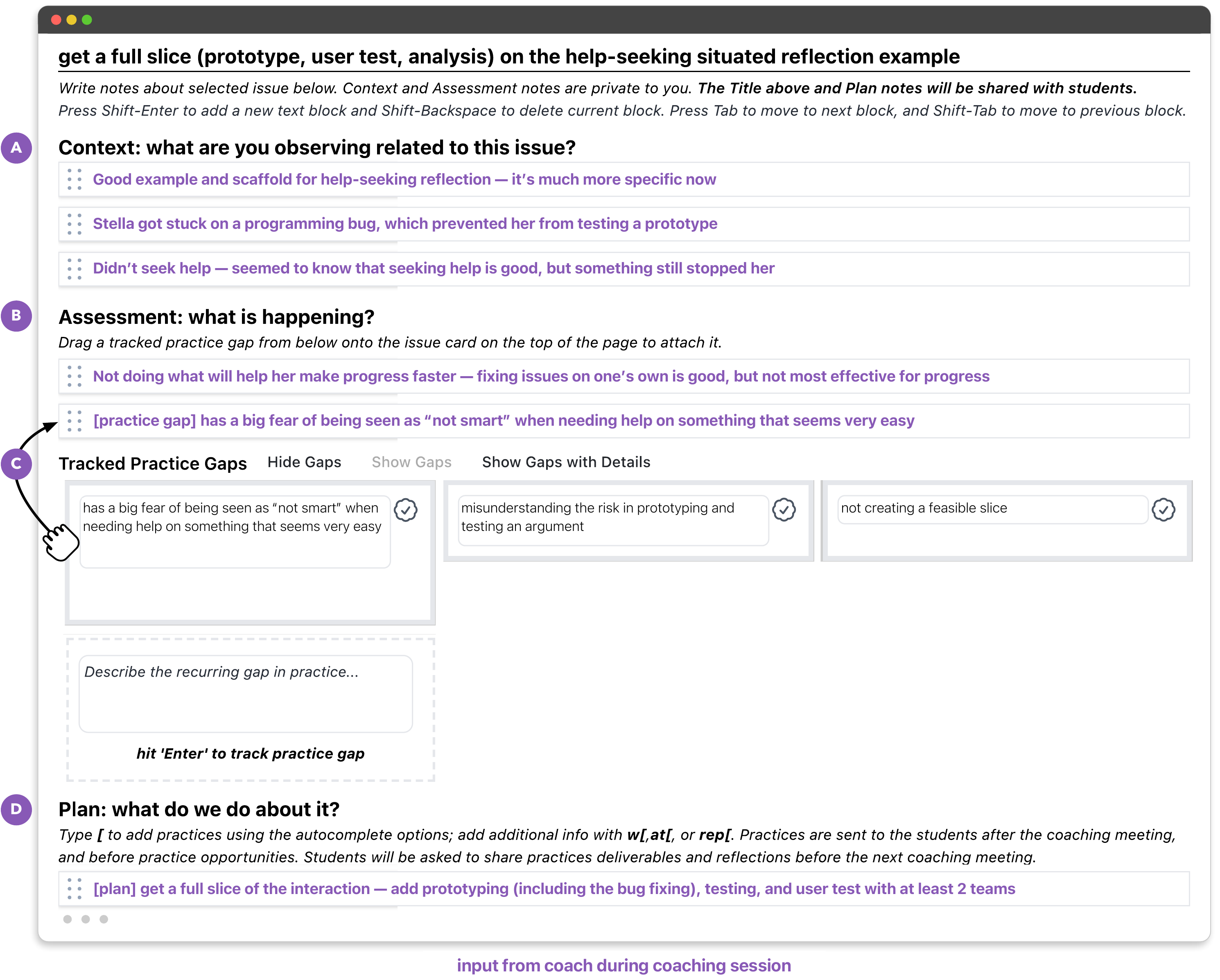}
  \caption{During and after the meeting, a coach uses Context-Assessment-Plan (CAP) structure to model a practice: the links between work issues, practice issues, and practice activities. First, they note what they observed in Context (A). Then, they consider what practice issues might be happening and why in Assessment (B). While conducting the Assessment, the coach can determine whether a recurring Tracked Practice Gap causes the current work issue and attach it to the Assessment by dragging the gap box; alternatively, they can track a new recurring gap if one is noticed (C). Finally, the coach suggests practice activities in Plan (D).}
  \label{fig:cap-notes-modeling}
  \Description[User interface of the Interactive Context-Assessment-Plan (CAP) Notes tool for modeling a practice.]{The interface has three sections. At the very top is the title of the issue for which the practice is being modeled. Below, the first section, Context, records observations about the issue. Next is the Assessment section, which records reasons for the work issue. This section also includes a Tracked Practice Gaps subsection, which displays reusable practice gap cards that can be attached to the Assessment by dragging and dropping. Finally, the Plan section outlines suggested practice activities.}
\end{figure}

During and after the meeting, CAP Notes provides the coach with a {\em Context-Assessment-Plan structure} to model links between work issues, (recurring) practice gaps, and alternative practices to try; see Figure~\ref{fig:cap-notes-modeling}. The structure helps the coach draw connections between practice issues, regulation skill gaps, and current work issues on a {\em per-issue basis}. For example, for a design project, the coach wanted the student to prototype and test a new system, but the student got stuck on a bug (Context; A). As the coach models the current work situation, they can consider if the issue relates to a recurring {\em Tracked Practice Gap} (e.g., not creating feasible plans). Here, the student shared that they did not seek help in reflections but, in further discussion with the coach, revealed a fear of being seen as ``not smart'', which the coach noted (Assessment; B). Coaches can also convert Assessments they recognize as recurring issues into a new Tracked Practice Gap that resurfaces in subsequent Assessment sections with Context from the specific work issues in which the gap occurred (C). Finally, the coach suggests practices for the student to attempt (Plan; D). By making Plans part of the structure for a specific issue, the coach can see whether suggested practices address the work issue {\em and} the underlying practice or regulation gaps identified in the Assessment section.

\subsection{Practice Agents: Facilitating Practices for Students}
\begin{figure}[t]
  \centering
  \includegraphics[width=0.75\textwidth]{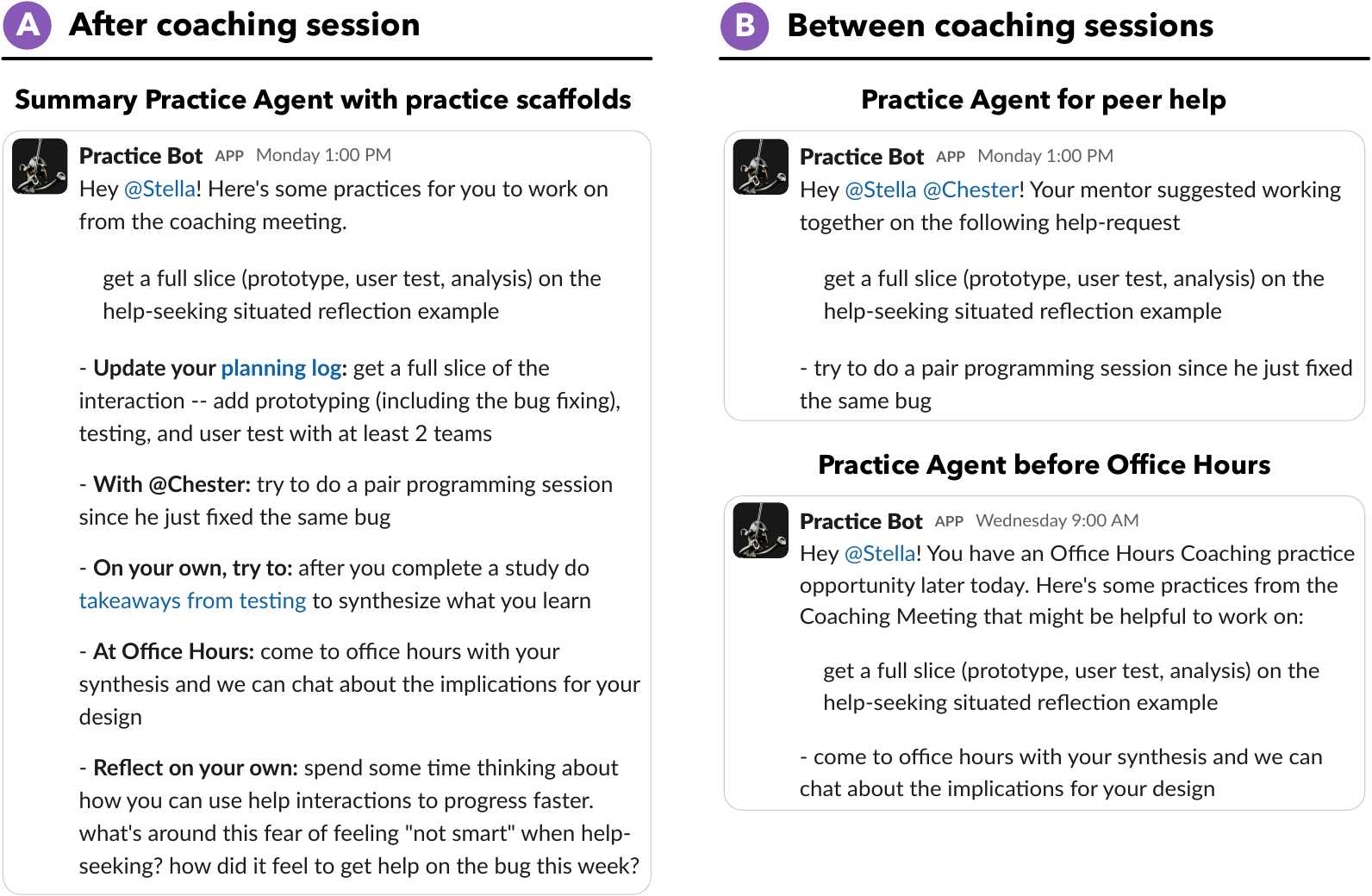}
  \caption{To facilitate practices, coaches describe Practice Agents in the Plan part of CAP Notes using a domain-specific language that includes the practice activity (e.g., re-planning; help-seeking; self-work; reflection on a specific prompt) and any resources (e.g., working structures, specific people or venues). These are compiled to include links to the student's specific tools or documents (in blue) and are presented via Slack after the coaching session (A). Practice Agents surface practice opportunities between coaching sessions, such as initiating a group message with a potential helper or a reminder about an upcoming Office Hours (B).}
  \label{fig:cap-notes-facilitating}
  \Description[Practice Agents delivering practices via Slack.]{Two panels showing Practice Agents at different times. Panel A, 'After coaching session', shows a Summary Practice Agent sending Stella, the student, all of the coach's suggested practices, with links to Stella's productivity tools. Panel B, 'Between coaching sessions', shows two follow-ups: a peer-help Practice Agent starting a message about working with her peer, Chester; and an office-hours Practice Agent reminding Stella to come to office hours to work with her mentor.}
\end{figure}

Coaches use a domain-specific language provided by CAP Notes in the Plan section to describe {\em Practice Agents} that correspond to core regulation activities for the student to practice:
\begin{enumerate}
  \item \lstinline|[plan]| suggests deliverables or tasks for the student to add to their planning log.
  \item \lstinline|[self-work]| captures a work activity for the student to attempt on their own.
  \item \lstinline|[help]| suggests an opportunity to work with a coach or peer for a work task. 
  \item \lstinline|[reflect]| prompts for reflections on a situation the coach observed.
\end{enumerate}
Coaches can optionally specify details about the practice, such as a representation or resource to use (e.g., \lstinline|rep[design argument]| for a design representation~\cite{lewis_planning_2018}), specific people (e.g., \lstinline|w[Jack]| \lstinline|w[Jill]|) to ask for help, or a specific venue (e.g., \lstinline|at[Office Hours]|) to practice in.

Following the coaching session, the student receives all of the coach's suggested practices---with references to specific tools and resources---via Slack; see Figure~\ref{fig:cap-notes-facilitating}a.\footnote{During testing, coaches did not want to share Context and Assessment notes with students since they included private observations. Practice Agents currently only share the issue title and compiled Plan notes with the student.} Between coaching sessions, Practice Agents follow up on the practices. In Figure~\ref{fig:cap-notes-facilitating}b, a \lstinline|[help]| agent initiates a group chat between Stella and Chester to resolve a programming bug; a separate \lstinline|[help]| agent reminds Stella to work on a practice with her coach the morning of the coach's Office Hours. If practices are not attempted, Practice Agents can remind students again. For example, the \lstinline|[plan]| agent will remind the student to re-plan if the planning log is not updated within 1 day of the coaching session.

\begin{figure}[t]
  \centering
  \includegraphics[width=\textwidth]{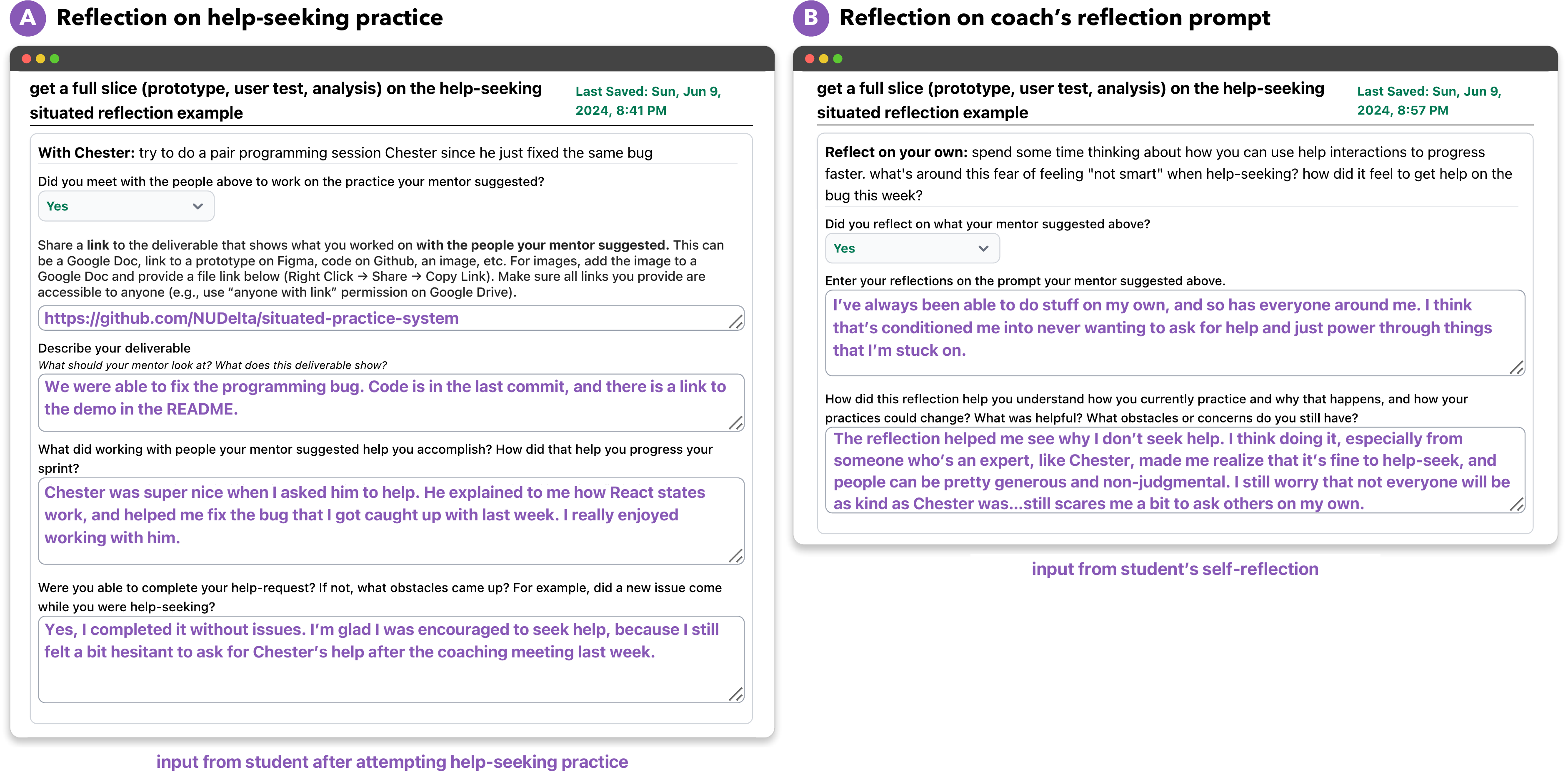}
  \caption{Practice Agents prompt students for practice-specific reflections before the next coaching meeting. In this example, Stella reflects on her help-seeking interaction with Chester, describing how it helped her fix the software bug, and on her regulation challenge of being seen as ``not smart'' when seeking help (A). Then, she reflects on why help-seeking is hard for her, a reflection prompt her coach provided (B).}
  \label{fig:cap-notes-reflection}
  \Description[User interface for students' post-practice reflections.]{Two side-by-side screenshots from the reflection tool. Each screenshot includes the following, from top to bottom: the issue title, the practice being reflected on, a drop-down indicating whether the practice was done, and text fields for practice-specific reflection questions. The left screenshot, 'Reflection on help-seeking practice', asks Stella to share a deliverable, describe it, explain how help-seeking advanced the work, and identify any challenges with help-seeking. The right screenshot, 'Reflection on coach's reflection prompt', asks for reflection on the mentor's prompt and how it can help Stella understand her current practice.}
\end{figure}

Before the next coaching session, the student is prompted with {\em practice-specific reflection scaffolds} in a reflection tool. In Figure~\ref{fig:cap-notes-reflection}a, Stella is asked to reflect on her help-seeking interaction with Chester, including sharing a deliverable and how it helped progress her work; in Figure~\ref{fig:cap-notes-reflection}b, she is asked to reflect on her coach's prompt about help-seeking. See Appendix~\ref{appendix:agent-questions} for reflection scaffolds.

\subsection{Tracking Regulation with Practice Objects and Creating Practice Scripts}
\subsubsection{Practice Objects for Linking Work Issues, Practice Gaps, and New Practices}
\begin{figure}[t]
  \centering
  \includegraphics[width=\textwidth]{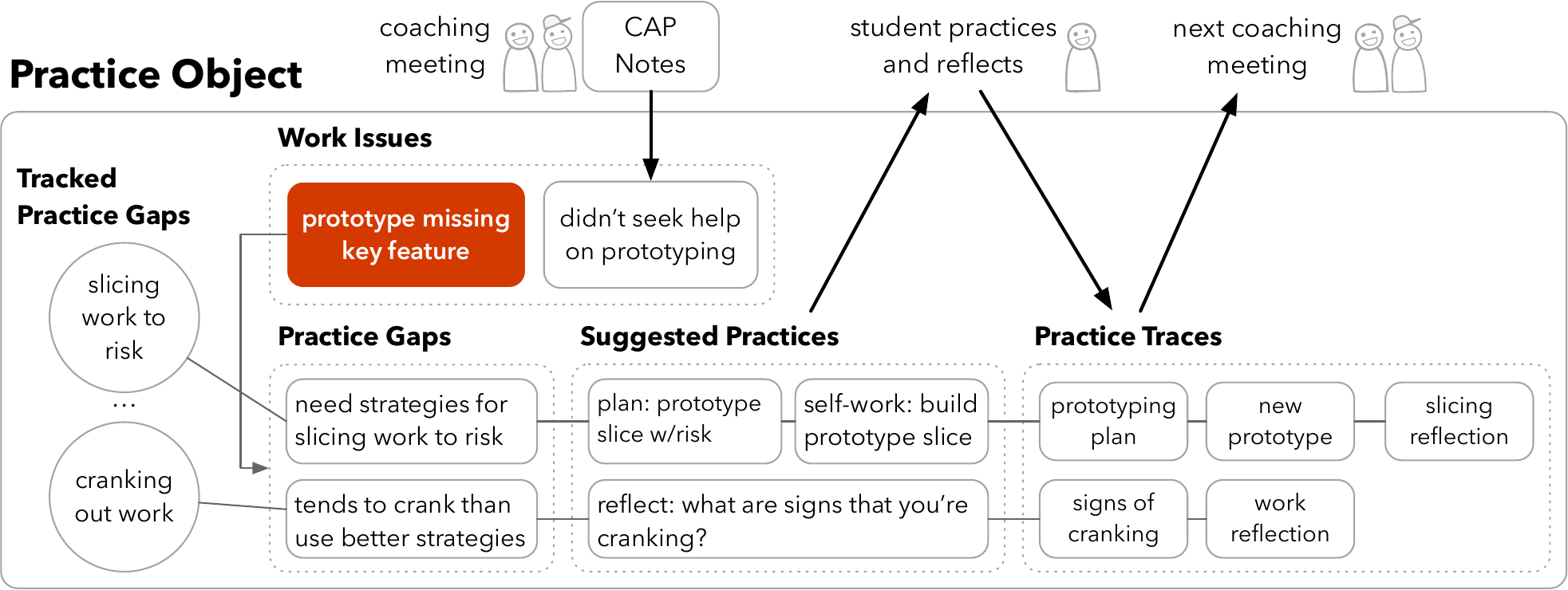}
  \caption{Practice Objects are an evolving representation of a student's work issues and regulation practices. A Practice Object is updated from coaches' use of CAP Notes during coaching meetings---storing current work issues, new Practice Gaps and Tracked Practice Gaps, and suggested practices---and from students' practice and reflections. CAP Notes resurfaces collected Practice Traces at the next coaching session for review.}
  \label{fig:cap-notes-practice-object}
  \Description[Practice Object data model linking work and practice information over time.]{A data model and lifecycle diagram inside one large Practice Object container. Work issues are added from CAP Notes. One work issue, 'prototype missing key feature,' is selected, for which two Tracked Practice Gaps, 'slicing work to risk' and 'cranking out work,' are attached. Each Practice Gap is in its own row and is attached to Suggested Practices. The student practices and reflects on these, with outcomes stored in Practice Traces, which are also connected to their respective Suggested Practices. Practice Traces are shown to the coach at the next coaching meeting.}
\end{figure}

To computationally represent students' work practice, we introduce {\bf Practice Objects}, a computational abstraction for maintaining and linking information about a student's evolving work issues, practice gaps, practice suggestions, and Practice Traces across coaching sessions and practice activities; see Figure~\ref{fig:cap-notes-practice-object}. The Practice Object is updated through CAP Notes during a coaching session. As practices unfold, traces of students' practice (activities, outputs, and reflections) are automatically added to the Practice Object to facilitate subsequent coaching sessions. Over time, a student's Practice Object effectively contains an interconnected history of the work issues they encountered, their coach's assessments of their practice gaps, and their regulation behaviors aimed at addressing those issues and gaps. 

\subsubsection{Practice Scripts for In-Situ Practice Support}
\label{sssec:practice-scripts}
\begin{figure}[t]
  \centering
  \includegraphics[width=\textwidth]{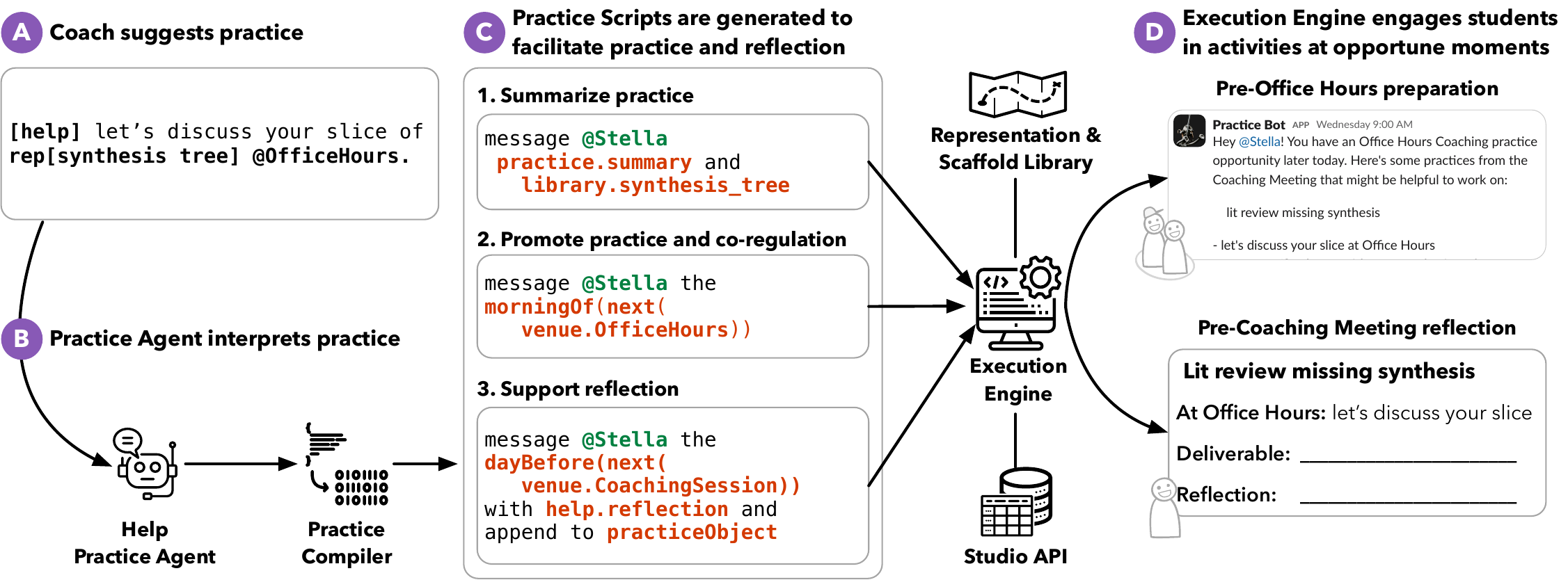}
  \caption{Practice Scripts enable a coach's suggested practice to be tracked and enacted for a student at the appropriate times. The coach's practice (A) is handed off to a Practice Agent, which interprets it via the Practice Compiler (B) to generate Practice Scripts. These scripts provide a practice summary, facilitate the practice, and facilitate reflection prior to the next coaching session (C). An execution engine based on Orchestration Scripts~\cite{garg_orchestraton-scripts_2023} tracks the Practice Scripts to engage students in practice and reflection activities (D).}
  \label{fig:cap-notes-practice-agent}
  \Description[Process of converting coach-authored practice into executed Practice Script.]{A four-part pipeline. Panel A shows a coach-authored help-seeking practice that asks the student to attend office hours with a synthesis-tree representation to discuss the implications for a design. Panel B shows a Help Practice Agent and Practice Compiler interpreting the practice. Panel C shows three generated Practice Scripts: one summarizes the practice; one surfaces the practice at the correct opportunity; and one supports reflection the day before the next coaching session. Panel D shows the Execution Engine using the Representation and Scaffold Library and Studio API to send a pre-office-hours practice reminder through the Slack Practice Bot and a pre-coaching reflection prompt for the same practice.}
\end{figure}

To make Practice Agents executable, we develop a \textbf{Practice Compiler} that generates \textbf{Practice Scripts} that surface practices at opportune times; see Figure~\ref{fig:cap-notes-practice-agent}. For example, when a coach uses CAP Notes to suggest that a student develop a synthesis tree to understand related work and bring it to Office Hours (A), it is handed off to a Practice Agent and interpreted by our Practice Compiler (B). This compiler extends \citeauthor{garg_orchestraton-scripts_2023}'s prior work on Orchestration Scripts~\cite{garg_orchestraton-scripts_2023}, which provide {\em Organizational Objects} computational abstractions that model the ways of working in an organization (e.g., processes; meeting venues; tools) and the {\em Studio API} that provides a unified database that maps abstractions to concrete data to make them executable (e.g., Team A meets on Monday at 10:00 AM); see Appendix~\ref{appendix:os-diagram} for details. The Practice Compiler creates a suite of Practice Scripts on-the-fly (C) that (1) summarizes the practice and shares relevant scaffolds with students; (2) reminds students of relevant practice and co-regulation opportunities; and (3) sends practice-specific reflection scaffolds to students before the next coaching session. Specifically, it uses a regular-expression-based parser that maps each Practice Agent to a script template and the relevant Organizational Objects. For example, in Figure~\ref{fig:cap-notes-practice-agent}, the \lstinline|[help]| plan specifies a venue with \lstinline|at[OfficeHours]|, which the parser interprets as the morning of the next Office Hours venue. Similarly, the parser injects links to matched representations specified with \lstinline|rep[]|, such as the synthesis tree. Finally, the {\em Execution Engine} tracks the execution of generated Practice Scripts across practice and reflection opportunities (D)~\cite{garg_orchestraton-scripts_2023}. As a student's practice progresses, Practice Agents will automatically collect their work and reflection and incorporate them into the student's Practice Object for review at the next coaching session.

\subsection{Design Iterations}
We made the following iterations to our design arguments, alongside addressing usability issues.

\subsubsection{Useful Data for Assessing Practices}
Early versions of CAP Notes did not include Practice Traces. CAP Notes initially displayed data from workplace tools (e.g., hours spent and main stories from a planning log) and a summary of prior practices at the top of the tool, but not linked to specific work issues, practices, or students' reflections and deliverables. Consequently, coaches had few signals to evaluate how well the practices they suggested went. We resolved this by having Practice Agents scaffold practice-specific reflections before a coaching session and include relevant tool data (e.g., updates to planning log), which composed the Practice Trace shown to coaches, providing {\em clearer signals of how students were practicing to generate work outcomes}.

\subsubsection{Loose Structure Before Rigid Structure}
CAP Notes' structure was initially too rigid, as coaches had to complete the Context-Assessment-Plan for each issue. We thought this would encourage coaches to consider practice gaps, but saw they needed to capture signals of issues {\em first} before modeling the practice. We introduced {\em loose structures} called Issues of Concern in which coaches could note their observations without a formal structure. When coaches are ready, these instantly transform into a complete Context-Assessment-Plan note by clicking the issue, thereby promoting formal modeling of the practice once issues are clearer. 

\subsubsection{Tracking Within and Across Weeks}
CAP Notes initially lacked Tracked Practice Gaps, making it harder to identify if a current work issue was caused by a recurring practice gap. We added these in the Assessment section, allowing coaches to {\em draw connections between recurring issues and the current work situation.} Moreover, the Practice Object stores the context of the current issue if a recurring gap is attached or if an Assessment is tracked as a new gap, supporting future diagnoses.

\subsubsection{Practice-Focused Versus Task-Focused Follow-Ups}
CAP Notes initially encouraged coaches to suggest task-oriented follow-ups, but not to consider regulation skills. For example, a coach could write, ``bring the bug to \lstinline|at[PeerHelp]| to get help'', but leave the practice implicit---a peer can help address code issues. We revised the Practice Agent's interaction to describe the {\em practice activity} a student could engage in, rather than the tasks to be done. For example, the coach would instead write ``\lstinline|[help]| get help from a peer on the bug,'' with the option to specify ``\lstinline|at[PeerHelp]|'' to further scaffold the student. This change enables coaches to explicitly outline the types of practices they want students to attempt, while providing sufficient scaffolding to support their efforts.

\subsection{Technical Implementation}
We built Situated Practice Systems using Next.js, a React framework for full-stack web applications, with TypeScript, Tailwind CSS, and MongoDB. CAP Notes creates a new note interface for each project each week, automatically fetching Practice Traces and Tracked Practice Gaps. We optimized the notetaking interface for laptop use during coaching. The system automatically saves edited notes to the MongoDB database. Practice Agents use the Slack API to send practice suggestions via the Practice Bot (Figure~\ref{fig:cap-notes-facilitating}). Our extended Execution Engine from Orchestration Scripts~\cite{garg_orchestraton-scripts_2023} tracks and surfaces practices in relevant opportunities, and prompts for practice reflections.

\section{Formative Field Study Methods}
\label{sec:formative-study-methods}
To understand how the latest iteration of Situated Practice Systems supports regulation-informed practice where earlier versions could not, we conducted a 3-week field study. We asked:
\begin{enumerate}[label=\textbf{RQ\arabic*:}] 
  \item How does CAP Notes help coaches observe and understand how students currently practice?
  \item How does CAP Notes help coaches assess practice issues and model new practices?
  \item How do Practice Agents help students facilitate regulation-informed practices?
\end{enumerate}

\subsection{Participants}
We conducted our study in the same research learning community described in Section~\ref{sec:dbr-method-setting}. Participants included two expert faculty coaches and seven student researchers (five undergraduates, one master's student, and one PhD student). For privacy, we use pseudonyms: Corey and Danielle refer to the coaches, and all other names refer to student researchers. Each faculty coach had several years of coaching experience. Students were members of the community for 1--3 academic terms (Mean = 2) and were in five project teams: two teams of two students and three individual projects. Corey advised all projects, while Danielle co-advised one project.

Because CAP Notes was used during the needfinding study, the system already had Tracked Practice Gaps for the students. Corey had provided feedback on CAP Notes during development and was familiar with the tool; Danielle first used it during the field study after an instructional session with the lead author. The lead author also introduced students to Practice Agents and explained that they would receive practice suggestions and reflection reminders via Slack. Corey was out of town for the first two weeks of the study and provided practice feedback asynchronously using CAP Notes, without a coaching meeting; Danielle led the coaching sessions of the co-advised project during the 3-week study.

\subsection{Procedure}
We instructed coaches to review students' deliverables and reflections in CAP Notes before each coaching session and use CAP Notes during coaching sessions to note observed issues and suggested practices for their students. Students received these practices one hour after their coach completed their notes. Practice Bot also sent practice reminders throughout the week; we instructed students to work as usual and use them when helpful. One day before their coaching session, students received a reflection tool to complete by the following morning.

During the deployment, coaches participated in weekly 20--30 minute semi-structured interviews in which they reviewed their use of CAP Notes that week. Students completed a 10-minute retrospective weekly survey after their coaching session. The survey included 5-point Likert-type questions (1: Strongly Disagree; 3: Neither Agree nor Disagree; 5: Strongly Agree) and open-ended questions about: (1) coaches' suggested practices; (2) the prompts for those practices at relevant opportunities; and (3) the reflection process on research progress and their practice. 

Following the deployment, we conducted semi-structured interviews. We asked coaches how CAP Notes helped them understand students' evolving practice issues over the weeks, how they reasoned about the practices they modeled and shared, and how their students' work processes and regulation behaviors changed during the study. We asked students how receiving their coach's practice suggestions after coaching sessions and throughout the week shaped their work, how practice-specific reflections influenced their understanding of their regulation gaps, and how their practices changed as a result of the tool. All participants were also asked to reflect on their concerns about using the tool. To aid recall, we showed coaches their notes for each student at each coaching session; similarly, we showed students summaries of their coach's practice suggestions. Post-deployment interviews lasted 25--40 minutes.\footnote{There were a few small data issues and study adaptations to report during the field study. Kiran's notes had a software bug during Week 2 that prevented his coach, Corey, from taking notes, resulting in Kiran not receiving any practices or a survey to complete that week. Will did not complete the Week 3 survey or post-deployment interview. Because Corey was out of town during the first two weeks of the study, we conducted only one interview during that period.}

Before the study, all participants provided informed consent and agreed to have their data used anonymously for research purposes. Before interviews, participants consented to audio, video, and screen recordings; we transcribed audio recordings for analysis. The lead author conducted all interviews in person or via Zoom. Student researchers were compensated \$20 USD for completing the weekly surveys and final interview ($\approx$1 hour total), but not for weekly reflections or following CAP Notes practice suggestions, which were considered part of their normal research work.\footnote{Will was provided partial compensation for his time in the study, namely \$10 for completing the first two weekly surveys.}

\subsection{Measures and Analysis}

Our analysis is grounded in two data sources. First, we analyzed CAP Notes system logs, Practice Traces, and student reflections to characterize: (1) the quantity and types of regulation gaps that coaches associated with identified work issues; (2) the quantity and types of practice suggestions offered; (3) the types of Practice Agents coaches used; (4) the frequency of Practice Agent follow-ups after coaching meetings; and (5) how often students attempted suggested practices. We also qualitatively coded coaches' identified work issues for cognitive, metacognitive, and emotional regulation gaps to understand the frequency of each gap; see Appendix~\ref{appendix:regulation-gap-coding} for our codebook. We similarly coded the content of coaches' practice suggestions. For example, we coded a practice suggestion that involved working with a representation as ``Representing problem and solution space''. The lead author and two undergraduate research assistants performed the qualitative coding.

Second, we analyzed open-ended interview and survey responses from coaches and students using thematic analysis~\cite{braun_using_2006}, focusing on three dimensions that aligned with our design goals:
\begin{enumerate}[label=\textbf{RQ\arabic*:}] 
  \item \textbf{Observing existing practices and potential practice gaps.} We coded coaches' interviews to understand how they used CAP Notes to understand prior practices. We coded students' survey responses and interviews to understand how reflections helped them communicate practice outcomes and self-perceptions of their practice.

  \item \textbf{Assessing practice gaps and modeling regulation-informed practices.} We coded coaches' interviews for how CAP Notes supported them in understanding regulation issues underlying work issues---including noticing recurring practice gaps---and modeling new practices. We coded students' survey responses and interviews to determine how they felt the suggested practices helped them regulate their work process during the week.

  \item \textbf{Facilitating regulation-informed practices.} We coded coaches' perceptions of their students' enactment of suggested practices and regulation skills. We coded students' survey responses and interviews to understand how Practice Agents supported their research practice throughout the week, and their reflections on regulation challenges.
\end{enumerate}
Across all research questions, we also coded for any challenges or concerns. After coding, we reviewed sub-themes generated from the analysis. We examined how the data confirmed or contradicted themes and refined them by splitting and combining themes until they were distinct; the final themes are presented below. The lead author conducted qualitative coding, initial theme generation, and initial theme review for the survey and interview data. All authors collectively refined and finalized themes. We also report quantitative data on students' responses to the weekly Likert-type survey questions. Together, these analyses provide a holistic account of how a Situated Practice System can develop and facilitate regulation-informed practices, using quantitative analyses when available and qualitative analyses of its impact on coaches' coaching and students' practice.

\section{Formative Field Study Results}
\label{sec:formative-study}

\begin{figure}[t]
  \centering
  \includegraphics[width=\linewidth]{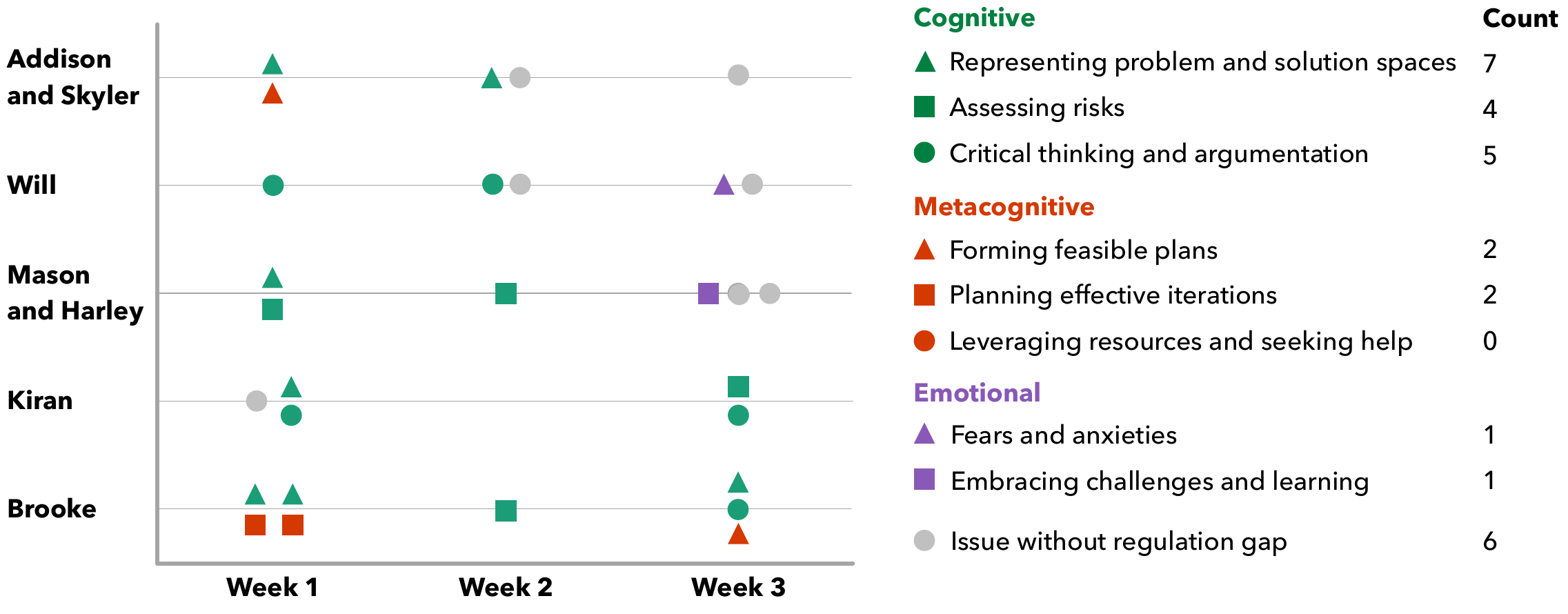}
  \caption{Across 21 work issues, coaches noted 22 regulation gaps across cognitive, metacognitive, and emotional regulation skills. Students could have multiple work issues (separated horizontally) with multiple regulation gaps (separated vertically). For example, in Week 1, Addison and Skyler had 1 work issue with cognitive and metacognitive gaps; in Week 2, they had 2 work issues: 1 with a cognitive gap and 1 with no gaps.}
  \label{fig:regulation-gaps-issues}
  \Description[Dot chart of work issues and regulation gaps by team and week.]{Dot chart with 5 rows corresponding to student teams and 3 columns corresponding to weeks 1 through 3 in the study. Dots represent which regulation gaps were present in each work issue. Addison and Skyler: Week 1, issue 1 had gaps in representing problem and solution spaces and forming feasible plans; Week 2, issue 1 had a gap in representing problem and solution spaces, and issue 2 had no regulation gap; Week 3, issue 1 had no regulation gap. Will: Week 1, issue 1 had a gap in critical thinking and argumentation; Week 2, issue 1 had a gap in critical thinking and argumentation, and issue 2 had no regulation gap; Week 3, issue 1 had a gap in fears and anxieties, and issue 2 had no regulation gap. Mason and Harley: Week 1, issue 1 had gaps in representing problem and solution spaces and assessing risks; Week 2, issue 1 had a gap in assessing risks; Week 3, issue 1 had a gap in embracing challenges and learning, while issues 2 and 3 had no regulation gaps. Kiran: Week 1, issue 1 had no regulation gaps, and issue 2 had gaps in representing problem and solution spaces and in critical thinking and argumentation; Week 2 had no work issues shown; Week 3, issue 1 had gaps in assessing risks and in critical thinking and argumentation. Brooke: Week 1, issues 1 and 2 each had gaps in representing problem and solution spaces and planning effective iterations; Week 2, issue 1 had a gap in assessing risks; Week 3, issue 1 had gaps in representing problem and solution spaces, critical thinking and argumentation, and forming feasible plans.}
\end{figure}

During the 3-week study, coaches identified and tracked 21 work issues among students. Across these issues, coaches identified 22 regulation gaps (see Figure~\ref{fig:regulation-gaps-issues}), 5 of which corresponded to previously identified Tracked Practice Gaps (see Table~\ref{tab:study-tracked-gaps}). While most gaps were cognitive (16), coaches also identified metacognitive (4) and emotional regulation gaps (2) that prevented students from engaging in relevant practices. Coaches did not add new Tracked Practice Gaps during the study. We hypothesize that because coaches had used the tool for almost 4 months, they had already identified students' recurring practice gaps by the time the study began. 

To resolve work issues and regulation gaps, coaches suggested 46 practices across the four Practice Agents (\lstinline|[plan]|: 8, \lstinline|[help]|: 5, \lstinline|[self-work]|: 29, \lstinline|[reflect]|: 4). Practice Agents shared practice summaries with students 14 times after coaching sessions. They also followed up 3 times for the 8 \lstinline|[plan]| agents, prompted students the morning of a coach's Office Hours for 2 \lstinline|[help]| agents, and started 2 Slack group messages to support peer help-seeking for 2 other \lstinline|[help]| agents.

In what follows, we describe how Situated Practice Systems supported regulation-informed practice by helping coaches observe current practices and potential gaps, assess why gaps occurred, and suggest new practices; we also describe how Practice Agents helped students enact those practices on their own and with others outside coaching sessions.

\subsection{RQ1: CAP Notes Helps Coaches More Easily Observe and Track Practice-Related Aspects of Students' Work}
\label{ssec:tracking-practice-dev}
\subsubsection{Understanding Emerging Practice Issues From Practice Traces}

By making work practices and regulation behaviors computationally representable, Practice Traces made students' practices and potential regulation gaps more apparent, even before the coaching session. Both coaches felt this helped their coaching. For example, Corey's students, Mason and Harley, were avoiding some problems when solutions were unclear by jumping to others. In their reflections, they described a possible direction as infeasible, which signaled to Corey that an underlying regulation gap might be preventing them from persisting. Corey said, ``In my mind, that direction is possible. So I know already [from the deliverables and reflections] that they have certain fixations or mental blocks that suggest they are worried about or unwilling [to think about that direction] or do not have good strategies [to work on it].'' This Practice Trace allowed Corey to focus the coaching session on the underlying regulation gap, rather than only on possible solutions to the immediate work issue. 

Students also found structured reflections valuable for understanding and communicating their research progress to their coach (Mean = 4.32; Agree to Strongly Agree), which promoted dialogue about practice issues. For example, Brooke shared how she was unclear about her takeaways from a recent user test: ``It was really helpful to go into that coaching meeting and be like, `all stuff was good, but here is this thing that I was weirded out by,' and be able to talk about that specifically.'' Contrasting CAP Notes with her earlier reflection practices, Harley shared: ``earlier, we were like, `oh, we hit this blocker,' but we did not tend to say, `what exactly is the blocker?'...With the reflection of [`what prevents you from doing something?'], we are more likely to take time and think through the list of obstacles. And that's something in addition to the original deliverable we would bring into the coaching meeting.'' In short, structured reflections helped students become more aware of work outcomes, obstacles, the practices that produced them, and ways to improve their practice.

\subsubsection{Loose Structures Help Coaches Track Fragments of Potential Issues to Discuss}

\begin{figure}[t]
  \centering
  \includegraphics[width=\linewidth]{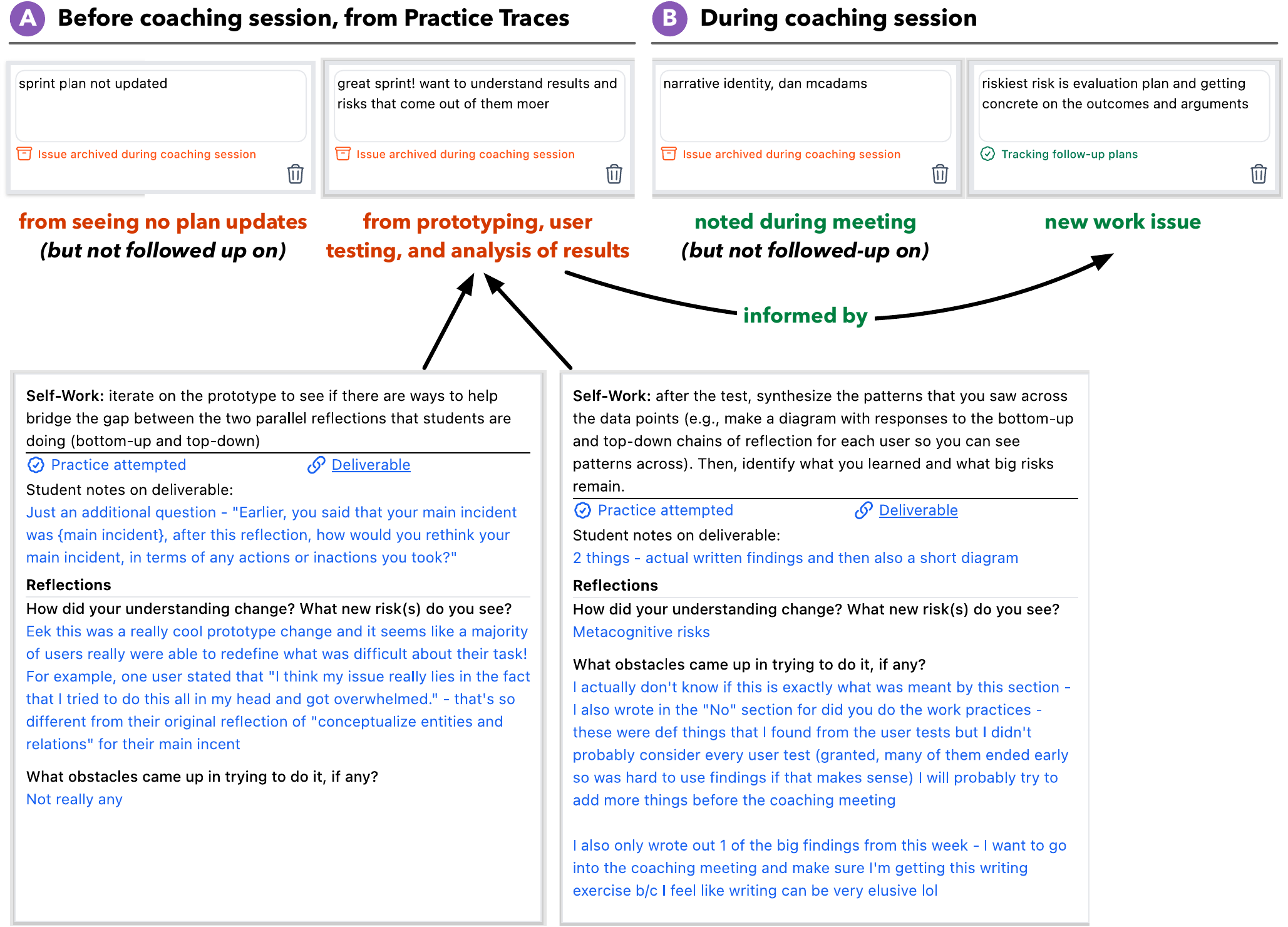}
  \caption{Not all issues identified from Practice Traces or conversations during coaching sessions were addressed with follow-up plans. For example, in Week 3, while working with Brooke, Danielle used Issues of Concern to note planning-related issues and early results from a prototype test to discuss after reviewing Practice Traces (A). During the coaching session (B), a relevant paper was mentioned and noted. However, she decided to focus on helping Brooke develop and execute a good evaluation plan, rather than reading more related work.}
  \label{fig:study-issues-of-concern}
  \Description[Annotated Issues of Concern noted before and during the coaching session.]{A two-panel annotated CAP Notes example. Panel A, 'Before coaching session, from Practice Traces', shows Issues of Concern noted during Practice Traces review, including an unupdated sprint plan and results from prototyping, user testing, and analysis. Panel B, 'During coaching session', shows two Issues of Concern noted during the coaching session: a narrative identity paper and a new work issue regarding testing on target users. The latter was informed by the prototype results noted in the earlier Issue of Concern.}
\end{figure}

CAP Notes provided coaches with loose structures for tracking signals of work and regulation issues that surfaced while reviewing Practice Traces and during coaching conversations. In Week 1, for example, Danielle shared that she, ``wouldn't have known that [Brooke] hadn't planned in advance [of the coaching meeting], and I wouldn't have thought to follow up with her,'' had the Practice Trace not surfaced Brooke's empty planning log and allowed Danielle to note it as an Issue of Concern to address in a Plan later on. In Week 3, Danielle chose not to suggest some literature that came up in conversation because she, ``didn't think that Brooke should go down a rabbit hole [right now]. So I wrote it down for myself, but I decided not to put that in the student-facing notes''; see Figure~\ref{fig:study-issues-of-concern}. In this way, loose structures helped coaches capture important observations while deferring decisions about whether and how to act on them until they had a fuller view of the practice issues.

\subsection{RQ2: CAP Notes Helped Coaches Assess Regulation Gaps and Suggest Regulation-Informed Practices}
\label{ssec:linking-issues-regulation}

\begin{table}[t]
  \small
  \centering
  \caption{Tracked Practice Gaps observed by coaches Danielle and Corey in students' work issues. Students not listed below had no work issues with prior practice gaps; no new gaps were created during the study.}
  \label{tab:study-tracked-gaps}
  \setlength{\tabcolsep}{3pt}
  \begin{tabular}{@{}
    p{0.10\linewidth}
    p{0.23\linewidth}
    p{0.23\linewidth}
    p{0.15\linewidth}
    p{0.23\linewidth}@{}}
    \toprule
    \begingroup\raggedright \textbf{Student(s)}\par\endgroup                                                                                                 &            
    \begingroup\raggedright \textbf{Tracked Practice Gap}\par\endgroup                                                                                       &            
    \begingroup\raggedright \textbf{Week 1 Issues}\par\endgroup                                                                                              &            
    \begingroup\raggedright \textbf{Week 2 Issues}\par\endgroup                                                                                              &            
    \begingroup\raggedright \textbf{Week 3 Issues}\par\endgroup \\
    \midrule

    \begingroup\raggedright Will\par\endgroup                                                                                                                &            
    \begingroup\raggedright Has maybe a fear of imperfection and may shy away from really doing the work where it's important but uncomfortable\par\endgroup & {\em None} 
    & {\em None} 
    &            
    \begingroup\raggedright Asking for references on digitization instead of trying it himself\par\endgroup \\
    \midrule

    \begingroup\raggedright Mason and Harley\par\endgroup                                                                                                    &            
    \begingroup\raggedright Plowing ahead / Not following through on a way of thought\par\endgroup                                                           & {\em None} 
    & {\em None} 
    &            
    \begingroup\raggedright Lacking insight on conceptual differences (and not writing that in the hypothesis) and not following through\par\endgroup \\
    \midrule

    \begingroup\raggedright Brooke\par\endgroup                                                                                                              &            
    \begingroup\raggedright Using good representations to guide thinking and writing\par\endgroup                                                            &            
    \begingroup\raggedright Diagram of your understanding of the relationship between beliefs, emotions, and actions
    \newline\newline
    Full slice (design, test, reflect) on the belief elicitation stage\par\endgroup                                                                          & {\em None} 
    &            
    \begingroup\raggedright Riskiest risk is the evaluation plan and getting concrete on the outcomes and arguments\par\endgroup \\
    \bottomrule
  \end{tabular}
  \Description[Table of tracked Practice Gaps recurring across students and weeks.]{A five-column table with rows for Will, Mason and Harley, and Brooke. Columns list the student or team, the Tracked Practice Gap, and work issues with the Tracked Practice Gap in Weeks 1, 2, and 3. Will and the Mason/Harley team have issues only in Week 3. Brooke has issues in Week 1 and Week 3. Week 2 is marked as none for all listed rows.}
\end{table}

\subsubsection{CAP Notes' Notetaking Structure Encouraged Coaches to Be Explicit About Regulation Gaps}

In 14 of the 15 coaching sessions, prompting coaches to consider regulation gaps while notetaking resulted in each project team having at least one work issue with an identified regulation gap each week, with 71\% of all work issues (15 of 21) having a regulation gap; see Figure~\ref{fig:regulation-gaps-issues}.

CAP Notes encouraged coaches to connect observed work issues to underlying regulation gaps. Although both coaches were already motivated to make these connections, they sometimes skipped that step and suggested practice based only on the work outcome. CAP Notes made this connection explicit in the notetaking process. For instance, Brooke struggled with planning effective research iterations and with using working representations to structure her analysis. In the Week 2 meeting, Danielle focused her notes on how Brooke's revised prototype should function when taking notes, but not on these regulation gaps. Upon review, she noticed a missing Assessment for the work issue: ``[CAP Notes] was really encouraging me to think in a different way about the student's work process.'' As a result, she noted that Brooke, `wasn't taking time to step back and think about her big takeaways and next steps based on what she learned,' as an Assessment. She then revised her Plans to suggest a corresponding practice, saying: ``I explicitly asked her to synthesize and told her a little bit of what I would want to see. I think she's struggling with this, and I reflected on that [in Assessment], which I might not have done if I were just writing free-form notes...I realized, `oh, yeah, I need to help scaffold [the synthesis] after the [user] test.' '' 

Because Practice Objects tracked practice issues over time, they enabled CAP Notes to help coaches recognize students' recurring regulation gaps across different work issues. Tracked Practice Gaps shown in the Assessment portion of CAP Notes helped coaches identify five new work issues that shared the same regulation issue as a prior practice gap; see Table~\ref{tab:study-tracked-gaps}. For example, Danielle shared how Brooke was, ``lacking good representation in her work and thinking'', saying, ``she comes in, and she's just spewing information that is not organized well.'' This recognition led Danielle to repeatedly suggest practices with representations to help Brooke structure her thinking. These cases illustrate how coaches can maintain and develop a model of students' practice across weeks that informs the practices they suggest, rather than only tracking task progress and troubleshooting. 

Lastly, CAP Notes helped coaches identify both easier-to-diagnose cognitive regulation gaps (e.g., using effective representations for thinking) and less apparent metacognitive and emotional regulation gaps (six of the 22 identified gaps). When a student struggles to progress, a coach's first reaction is often to suggest a cognitive scaffold (e.g., a diagram for writing in our needfinding study). However, when students fail to use these scaffolds, underlying non-cognitive regulation gaps often prevent progress. We found this was the case for all students except Brooke. For instance, Mason and Harley would quickly jump to new problems rather than persist with difficult ones. Corey initially suggested cognitive scaffolds to help them think more deeply (see Figure~\ref{fig:regulation-gaps-issues}, Weeks 1 and 2), but later realized that they were avoiding the problem when it became less clear and therefore not applying the scaffolds (see Figure~\ref{fig:regulation-gaps-issues}, Week 3). Will had a similar experience while writing a findings draft; Corey recognized from Will's past patterns that he was afraid to write about things he did not completely understand and confront the feeling of not knowing something (see Table~\ref{tab:study-tracked-gaps}), instead of ``speculating'' on unclear parts that might reveal new issues to address.

\subsubsection{Coaches' Practice Suggestions Embedded Regulation Strategies}

33 of the 46 suggested practices (71\%) included at least one regulation skill, suggesting that CAP Notes helped coaches connect strategies for progressing work to the regulation issues hindering progress. With CAP Notes, the links among work outcomes, practice issues, and new practices were explicit, as Corey explained: ``Plans were synthesized and structured: here's the [work] issue, here's some of the underlying [practice] reasons why it's happening, here's what we could do to address those underlying reasons. So that we address this issue, and we'll bring in this kind of deliverable.'' For instance, Addison and Skyler were struggling to plan effective milestones for their study analysis, particularly how to represent their findings. Corey could have suggested, ``plan out some milestones for analysis,'' but because he noticed that their existing representation was too high-level, he included: ``Can you think of a set of questions/goals you might have for understanding the example? Can you break it down more (your representation is still pretty high level---and not very detailed)?''

Students found that coaches' practices suggested clear ways to self-regulate while working (Mean = 4.06, Agree to Strongly Agree). Before CAP Notes, students struggled to capture the specific aspects of practice their coaches wanted them to try. For example, a coach might suggest where the risks in the students' research lie, but not specific scaffolds that could help (e.g., a study design template for planning a user study). Students could also forget the details of how to practice, when many issues and practices were discussed. Written practice suggestions helped students prioritize what their coach considered important and how to work on it. Mason said, ``Sometimes we went over many different problems in the research we are facing, and just remembering from the coaching meeting, sometimes we can miss some important points. And the practice [we get sent] is very directly pointed to what we should be focusing on this sprint....there was a suggestion about how to think about the distinctive difference and the features. That feedback from our coach is especially helpful to give us some guidance towards solving the problem a bit more by ourselves.''

\subsection{RQ3: Practice Agents Facilitate Students' Practice Through In-Situ Reminders and Post-Practice Reflections}
\label{ssec:practice-suggestions-action}
\subsubsection{Students Found Their Coach's Suggested Practice Actionable}

Practice Agents helped students enact their coaches' suggested practices. Based on students' reflections, 36 of the 46 suggested practices were attempted ($\approx$78\%), including 5 of 8 \lstinline|[plan]|, 23 of 29 \lstinline|[self-work]|, 4 of 5 \lstinline|[help]|, and 4 of 4 \lstinline|[reflect]| practices.\footnote{Will was suggested 2 \texttt{[self-work]} practices in Week 3 of the study. However, he did not complete the final survey, interview, or reflection tool. As such, we include him as not attempting the suggested practice here.}\textsuperscript{,}\footnote{In survey responses, students shared they did not attempt 3 cases of \texttt{[self-work]} due to limited time in the final week of the study, and one case due to running out of time for the practice that week. For the single \texttt{help} practice not attempted, the students met with their coaches elsewhere and did not need to schedule a separate meeting.} Students found that Practice Agent reminders helped them decide how to focus practice throughout the week (Mean = 4.06, Agree), and that the reminders were timely for practice opportunities (Mean = 4.25, Agree to Strongly Agree).

Before Practice Agents, students and coaches were often misaligned about how to practice. Danielle shared: ``[Before SPS], there were often miscommunications about what work should be done, where the student did stuff that wasn't in line with what we're looking for. That definitely didn't happen [during the study].'' By sharing coaches' detailed practice notes after the coaching meeting, Practice Agents clarified which issues to focus on and which regulation strategies to apply. Skyler shared how, ``there's a lot of information being thrown back and forth [during coaching], that makes it hard to have that, `Okay, here's exactly the next thing to do or what to keep in mind,' and I think it helps to receive that from the coach in a distilled version.''

Practice Agents helped students practice on their own and co-regulate with peers and coaches outside coaching sessions. One mechanism was initiating conversations, which lowered the barriers to reaching out to peers. For example, Mason, Harley, and Kiran---who worked on similar projects within the same research sub-group---all described how they missed opportunities to work together on a similar problem until one of Corey's \lstinline|[help]| agents reminded them to. Harley shared, ``sometimes you forget to follow up, especially with your group-mates...it broadens your thinking that way...[The reminder to set up a meeting] was needed...we had been pushing it off.'' Practice Agents also helped students prepare for upcoming help venues (e.g., Office Hours) rather than simply showing up, as they had often done before. Harley shared: ``[it was helpful] having that reminder if Corey suggests you come to Office Hours, then probably he thinks something is risky, and you should either talk about it ahead of time or prepare something to talk about.'' Skyler shared a similar experience: ``it was easy to not adequately prepare for Office Hours when sometimes things slip our mind as we go throughout the week.''

Practice Agents also increased coaches' confidence that students were engaging with their regulation-informed practice suggestions. Corey said: ``I have not had to worry about what students practice during the week, nearly as much as I used to. I used to check in on people mid-week, like, `hey, what did you take out of the coaching meeting? What are you practicing right now? Did you do this or that?', to keep tabs on all the students, and I don't have to anymore.'' Coaches also felt that students were progressing in line with their expectations. Danielle shared: ``[previously] things got lost in translation, and what ends up on their planning log, or what they end up doing, is not in line with what you thought they were gonna do. [With CAP Notes], it was like: here's the three things I want you to do, and [Brooke] did those three things.''

\subsubsection{Students Began to Understand How Their Regulation Skills Were Developing}

Every student described attempts to address regulation gaps that had hindered their research practice, as Practice Agents promoted reflection on those patterns (Mean = 4.11, Agree to Strongly Agree). In the final interview, Addison and Skyler described becoming more comfortable working with incomplete or imperfect understanding, a pattern that had affected their ability to apply Corey's suggested strategies. Addison shared, ``I've been trying to better understand that it's okay to produce something that's not good...I feel like I have, previously, and still am, been too focused on doing the best possible job to the point where it prevents me from working.'' Kiran shared that he had struggled to think deeply about problems because he would jump from topic to topic when a problem became difficult. In contrast, now he tries: ``to approach my risks more head-on, and stop avoiding my risks, and also be more patient and then think problems more thoroughly, give a little more time to process.'' Finally, Brooke described learning to use better representations for thinking and communicating her research, an issue Danielle had previously raised: ``[Before] I was fantastic at making things that [Danielle] did not understand...Now, I see representations as a way to tell a story about what I've learned (and share details through discussion).'' Corey felt that students advanced both their work and their regulation skills, sharing: ``It seems to me that [students] receive [the practical advice on the project and the link to some regulation gap]---like they actually got it and are practicing on regulation issues while they're working, or at least very aware of them now.'' These findings suggest that Situated Practice Systems can help students recognize how recurring mindsets, attitudes, and practice patterns hinder their research, and can support the development of regulation strategies or ways of working that make them more effective beyond the immediate work issues. 

\subsection{Potential Challenges With Situated Practice Systems}
While our study shows early promise for how Situated Practice Systems can support the development and enactment of regulation-informed practices, we also observed two notable challenges.

\subsubsection{Practice Trace Quality and Understanding Practices}
Although Practice Traces were critical for coaches' understanding of students' practices between coaching sessions, their quality varied. Unclear deliverables or poorer reflections provide weaker signals of potential regulation issues, requiring more diagnostic work during coaching meetings. Corey shared: ``you don't really get everything that students are thinking by just looking at the deliverables and a really brief reflection. When the reflections are better, you get a [more of a] glimpse into how the students are thinking about [their work and practice].'' Students also noted that producing high-quality deliverables and reflections can be challenging. For example, Harley said: ``we struggle with representing our deliverables in a way that makes sense even without you describing it.'' Mason found some reflections challenging, saying: ``I always struggled to answer, `How did your understanding change?' Sometimes, it's hard to identify the change in understanding we have.'' Thus, while Practice Traces made students' practices more legible to coaches, producing useful deliverables and reflections is itself a developing regulation skill that students may require coaching on.

\subsubsection{Level of Scaffolding Provided to Facilitate Practices}
Sharing practices with students requires balancing the scaffolding needed for enactment with the other regulation skills students need to practice. On the one hand, practices must be sufficiently scaffolded so students can enact them. While students generally found practices helpful, they sometimes did not feel they were actionable (Mean = 3.89, between Neither Agree nor Disagree and Agree), or it was unclear what successful practice looked like (Mean = 3.79). Sometimes, practices included high-level deliverables but few details. For example, Addison shared: ``one of the practices was `complete a full slice' [of needfinding, prototyping, and user testing] which was difficult given that we weren't completely sure what a full slice should look like.'' On the other hand, practices can be {\em too} scaffolded, taking away from the student the development of planning-related regulation skills. Early in the study, both coaches felt their guidance was too explicit and detracted from the planning work. Later, their suggestions became more open-ended, specifying goals while leaving room to plan details. Danielle shared: ``I'm leaning away from giving [Brooke] explicit direction on things that I want her to adapt based on how things are changing over time, versus the things that I put, which are all definite intermediate deliverables [to do].'' Harley described these as scaffolds for prioritizing: ``Especially with the suggestions of, `Come to Office Hours,' `work on this before the Lab Meeting,' even having a suggestion with that [opportunity], helps you think about planning your work during the week and what you should have done by when.'' Put another way, allowing students to fully self-direct their work risks missed opportunities to practice specific regulation skills. Conversely, when coaches structure those opportunities, they may bypass planning-related regulation skills. Understanding this balance and which skills require practice at a given time will be critical to providing students with practice across the range of skills needed to become independent practitioners.

\section{Discussion and Future Work}
There is a pressing need for college students to develop the regulation skills needed to lead innovation work, yet even expert coaches can struggle to help without computational support. We introduced Situated Practice Systems, which embed computation into the process of understanding, modeling, and facilitating regulation-informed practices for innovation work. Our user study demonstrates how our system supports coaches' understanding of and modeling of student practices, and facilitates students' practice of new strategies outside of coaching sessions. Below, we revisit the design principles of Situated Practice Systems, suggest future technical directions to improve their efficacy, and reflect on the design of future learning environments to support practice development.

\subsection{Design Principles of Situated Practice Systems}
\subsubsection{Focus on Practice and Regulation, Not Tasks}
Designing authentic learning environments for developing a practice and building regulation skills has historically been challenging because the coaching task is complex and difficult. A core innovation of our work is {\em developing tools that explicitly support understanding, modeling, and facilitating students' practice and regulation}. In our study, we found that these tools helped coaches and students to reflect on and discuss students' practices and regulation gaps. Coaching meetings were more practice-centric, and coaches were able to build an explicit model of how their students were practicing, understand {\em why} practices were failing, and suggest new {\em regulation-informed practices} that not only address issues with work tasks but help students to recognize and work with their underlying regulation gaps (see Figure~\ref{fig:regulation-gaps-issues}). Put differently, Situated Practice Systems facilitated coaches focusing on developing the regulation skills needed for students to self-direct complex work and to become independent (teach them how to fish), rather than to simply troubleshoot their projects (give them fish).

\subsubsection{Make Work Practices and Regulation Strategies Computational}
While highlighting regulation gaps and introducing new practices are helpful, students often need additional support to implement them. Situated Practice Systems {\em make work practices and regulation strategies computational}. This allows it to provide software agents that scaffold practice and reflection. Coaches can then provide tailored guidance on how the student should practice, which is then surfaced during practice opportunities outside of coaching meetings. In our study, we observed that coaches suggested 46 practices that provided concrete guidance and resources to help students practice outside coaching sessions. In this way, making work practices computational allows a coach to extend their support by facilitating the (suggested) use of diverse communal learning resources beyond the coach. 

\subsubsection{Support Working With Recurring Regulation Gaps and Patterns}
Gaps in regulation skills are often persistent issues for students that recur across work situations and tasks. Coaches can struggle to realize that the same pattern is affecting a new work issue, especially when tracking regulation gaps across many students. Situated Practice Systems {\em foreground recurring regulation gaps and encourage explicit reflection on them in connection to work issues}. Part of this tracking involves identifying how disparate work issues may share the same underlying regulation gaps, for which coaches can provide tailored practice strategies to address specific instances of these recurring gaps. In our study, our system helped coaches identify recurring regulation gaps for four of the students (see Table~\ref{tab:study-tracked-gaps}) and enabled them to suggest tailored, {\em situated} strategies for continued practice. This helps coaches view a student's struggle less as a set of disconnected work issues across weeks and more as an unfolding of the same recurring regulation gaps and patterns across situations, which they can help students learn to address. 

\subsubsection{Integrating Humans and Machines for Regulation Coaching}
Lastly, we designed Situated Practice Systems to involve humans and machines in complementary ways. Coaches are in control of the diagnosis process, while being supported by {\em computational affordances that track and surface relevant practice information.} Once a coach diagnoses a regulation gap and suggests a new practice, it is then handed off to a machine to facilitate. In contrast, AI systems that fully automate coaching (e.g., ITS in closed domains~\cite{alkhatlan2018intelligent, koedinger2006cognitive}; LLM-based coaches~\cite{huang2025ai,huang_intelligent_2023}) are not capable of diagnosing regulation gaps on their own, and may be undesirable anyway given the importance of building the student-coach relationship (see Section~\ref{ssec:disc-learning-envs}). Instead, our approach clearly delineates and uses the complementary strengths of coaches and computation to support diagnosis and practice.

\subsection{Future Directions for Situated Practice Systems and Regulation Coaching Tools}
\subsubsection{Practice Agents and an Ecosystem of Support}
Practice Agents open a rich interactional space for how software agents can support practice development. Our current implementation focused on a small set of core regulation skills (i.e., planning, self-work, help-seeking, and reflection) and provided simple scaffolds for practicing them (e.g., static templates), which already helped students engage in practices that address work issues and regulation gaps identified by coaches. It also allowed coaches to offload the need to present practices to students and to track them manually across the week. In the future, Practice Agents could enlist other systems that provide practice and task support, such as planning tools~\cite{shah_agile_2023, shah2026compass}, structured reflection tools~\cite{pribble_mindyoga_2022}, and help-finding tools~\cite{mcdonald_expertise_2000, ackerman_answer_1996}. Furthermore, Practice Agents could incorporate advances in large language models (LLMs), for example, by custom-tailoring practice or reflection scaffolds~\cite{huang2025ai}. 

\subsubsection{Practice Objects and Long-Term Development}

Practice Objects help to track a student's practice over an extended period of time. Students may need to develop many regulation skills to pursue innovation work, which can be challenging to do simultaneously. This requires coaches to select which to prioritize, for instance, at times under-emphasizing planning skills in favor of teaching representational structures for thinking. Moreover, seemingly ``resolved'' regulation gaps can resurface in new work issues, such as when a student who no longer fears writing discovers a fear of building systems. Future iterations of Situated Practice Systems could help coaches consider how to support student practice over longer time horizons, for instance, by considering which practices and skills are most developmentally supportive at any given time.

\subsubsection{Difficulties in Diagnosing Regulation Gaps}

Identifying regulation gaps can still be challenging, as their patterns may only become apparent over time and only after being recognized in some work issues. In our study, we observed that it took Corey 2--3 coaching sessions to realize why Will, Mason, and Harley could not fully apply the suggested cognitive scaffolds. While underlying regulation gaps will remain difficult to identify, some technologies may help. For instance, we envision extending the Assessment portion of CAP Notes with a {\em diagnostic environment} that surfaces common practice challenges across different tasks (e.g., scoping a complex prototype; managing emotional reactions when working on something new). In a different direction, we can advance the design of learning communities and practices that encourage students to proactively reflect on and share regulation-related challenges with coaches (see Section~\ref{sssec:disc-culture-srl}).

\subsubsection{Managing the Increased Work From Situated Practice Systems}

While Situated Practice Systems are beneficial for practice development, our study revealed that they increased the time spent on work for both students and coaches. On the student side, while much of the additional work---constructing clear deliverables and reflecting on their practice---is an important part of self-direction, coaches will need to set expectations for this work and create space for it as part of the learning curriculum (i.e., some work hours allocated to reflection and documentation) so it is seen as a learning activity and not ``busy work''. On the coach side, SPS adds additional work beyond the coaching session time they already provide, which can quickly become intractable if they want to use our system for coaching larger numbers of students. Such additional work can create an imbalance between those who put in the work and those who benefit (a historic CSCW challenge~\cite{grudin_why_1988}), affecting the long-term viability of using our system. In our study, both coaches shared that continued use would require small adaptations to the coaching meeting structure, allowing them to model practices within the meeting's time constraints (e.g., 25 minutes for mentoring, 5 minutes for modeling). Future work can also consider dispersing the responsibility of understanding and modeling practices to more senior peers in a learning community~\cite{garg_networked-orchestration_2022}, or consider the adoption of LLM-based tools to support modeling practices and regulation gaps (e.g.,~\cite{schimpf2026supporting, fang2026llmbased, ahn2026answer, zhu2026scaffolding, bhattacharjee2024understanding, edwards2024mai, huang2025ai}), including generating initial practice suggestions that coaches and students can then refine.

\subsection{Towards Practice-Oriented Learning Environments}
\label{ssec:disc-learning-envs}
While we argue that technology is needed to support the development of regulation-informed practice, questions remain about how to create effective learning environments for practice development in which such technology will succeed. To close, we reflect on our experience in the studied DTR community as lessons for future learning environment designers.

\subsubsection{Developing a Culture for Learning Regulation Skills}
\label{sssec:disc-culture-srl}
Situated Practice Systems require learning communities that foster a culture in which students want to move towards becoming independent innovators, which requires introspection into themselves and their regulation behaviors. In DTR, these are core learning objectives supported at every level, whereby goal-setting, coaching, and evaluations are entirely practice-centered and regulation-oriented instead of product or project outcome-focused. Moreover, the community makes time to discuss the challenges of self-directing research, spending up to an hour each week sitting in a circle to reflect together on how research is progressing, again with a focus on regulation challenges. Such activities normalize the challenging process of learning regulation skills and celebrate wins when a student shares how they are overcoming a long-standing regulation gap. In addition, addressing ineffective patterns and developing new self-beliefs takes time. While it may be tempting for coaches to scaffold students past their current level of development to accelerate project progress, we argue that communities that focus on learning to self-direct innovation work must value and prioritize developing the regulation skills needed for students to become independent, rather than prioritizing short-term productivity. 

\subsubsection{Shifts in Coach and Student Responsibilities}
Having a culture of regulation-informed practice shifts the roles of coaches and students within the community. In many PBL settings, students focus on developing domain-specific skills through an applied project, with the coach providing project-specific feedback. In a regulation-focused setting, students' focus shifts to understanding their practices and learning more effective regulation strategies, while the coaches' responsibility shifts to helping students grow in this way. For example, in DTR, coaching sessions were primarily focused on students reflecting on their work and exploring new ways to practice, with project feedback reserved for the end of the meeting. 

While we expect other communities will need to make similar shifts, they are likely to be uncomfortable. Students are asked to introspect and to work on letting go of (potentially long-standing) patterns in favor of more effective behaviors. Coaches, who are accustomed to understanding project issues, may feel uncomfortable or ill-equipped to ask students about their emotions and beliefs---for instance, their fears or anxieties---and may require additional training and preparation. Acknowledging such discomforts and training gaps, and recognizing the value of regulation-informed coaching and practice, will be crucial to adopting a practice-focused orientation in learning communities. 

Moreover, as AI tools for work and learning become increasingly capable in domain-specific skills (e.g., coding, writing), we believe the most effective practitioners will be the ones who practice, and who can effectively regulate their work and learning while incorporating AI tools as collaborative agents rather than fully offloading their work~\cite{tankelevitch2024metacognitive,tang2026how,ramesh2026metacognitive}. Such changes in how we work and learn further highlight the need for learning environments to emphasize the development of regulation skills over domain-specific skills alone.

\subsubsection{Technology-Inspired Changes in Coaching Practice}
Situated Practice Systems allow us to ask more of coaches: to look across work issues and skills, see how the student is progressing at the practice level, and prioritize addressing regulation gaps rather than only troubleshooting project progress. In other words, Situated Practice Systems help transform the {\em practice of coaching} from project-focused to regulation-focused by integrating computation into the coaching process at a level previously unexplored in coaching or productivity tools. Communities interested in practice and regulation will increasingly need to consider adopting practice-oriented tools such as SPS or updating their existing tools to be more practice-oriented. For example, software-focused communities could modify tools like GitHub to better facilitate practice-oriented discussions between experienced and novice developers throughout the planning and development process.

\subsubsection{Psychological Safety and Ethical Considerations}
\label{ssec:safety}

Lastly, all students need to feel psychologically safe to be open and transparent about their practice and regulation. Students may not feel comfortable sharing their struggles. They may fear judgment from peers or their coaches, or be understandably concerned that what they share could lead to a negative assessment during grading. To foster trust among students, coaches, and peers, we recommend the following. First, any regulation gaps shared by a student remain private to them and their coach unless they choose to share them more broadly, both in conversation and within Situated Practice Systems. Second, while openness is encouraged, students must retain control over whether and when they discuss a particular issue, without fear of negative consequences. Third, any formal assessments (i.e., grades) in the learning environment should reflect students' efforts and progress toward developing a regulation-informed practice, rather than whether they have reached a predetermined level of competence within a prescribed timeframe. Above all, fostering greater compassion for students' regulation challenges and who they are as people can help us develop learning environments in which students' current {\em and} future selves are both seen and held. We meet students where they are on their path toward becoming more independent and able to lead innovation work.

\subsection{Study Limitations}

While we focused on developing regulation skills, we did not assess whether students could independently apply them in new situations. Although some evidence suggests that students self-monitor for practice opportunities, our study's structure and length do not allow us to conclude that students can apply practices effectively and independently, especially given the challenges of applying abstract practices to new work situations. Future studies could examine how well students learn practices and how a Situated Practice System supports them.

Second, the research team's involvement in the community may bias our study. We chose this community because of the need for regulation skills to participate in the work (research), the emphasis that coaches place on {\em teaching} those skills, and students' desire to {\em learn} them. Moreover, this community already had rich tooling (e.g., planning, help-seeking, and communication tools with APIs) that our system could build upon, as well as numerous collaboration structures in which students can co-regulate practices with peers and coaches outside of coaching sessions. Future work can investigate these tools in other learning environments, including how our approach is applicable in other contexts and what adaptations are necessary. 

Finally, while our formative study already shows promise for how technology can support regulation skill development, future work should conduct further summative studies. These could include larger learning communities, other settings (e.g., entrepreneurship~\cite{huang_intelligent_2023}), and controlled comparisons, such as a basic notetaking tool without CAP Notes' structured notetaking affordances or Practice Agents that share summaries but not in-situ practice suggestions. These studies could also measure the difficulty of the regulation coaching task, including increased cognitive overhead. These would help reaffirm our formative results and provide further insights on how Situated Practice Systems support regulation-informed practice (e.g., whether practices are transferable).

\section{Conclusion}
Despite the critical need for students to lead innovation work, college project-based learning (PBL) environments risk underpreparing them by over-focusing on project progress rather than the core cognitive, metacognitive, and emotional regulation skills required for such work. While expert coaching can help, diagnosing and tracking students' regulation behaviors remain challenging without computational support. We introduced {\em Situated Practice Systems (SPS)} that use computation to support understanding, modeling, and facilitating regulation-informed practices through an {\em Interactive Context-Assessment-Plan (CAP) Notes} tool for coaches to understand and model students' practices and {\em Practice Agents} for students to help facilitate their practice. SPS uses {\em Practice Objects} to represent practices and regulation behaviors computationally and their evolution as the student practices, and {\em Practice Scripts} to automatically surface practice suggestions in relevant situations. Our formative 3-week field study in a research learning community demonstrates how SPS helped coaches provide regulation-informed practices and guided students in adopting new ways of working on their own and with others. These findings hold promise for the role that computation can play in shifting learning communities toward becoming more practice- and regulation-oriented for developing the next generation of innovators.

\begin{acks}

  We thank the Design Technology and Research (DTR) program at Northwestern University for participating in our study; Allyson Lee and Terry Chen for assistance with data analysis in the field study; Leesha Shah and Matt Easterday for helpful discussion on the conceptual framing; and Yinmiao Li, Melissa Chen, Evey Huang, the Delta Lab, and the CollabLab for their feedback. Research funding was provided by the \grantsponsor{DRL2506333}{U.S. National Science Foundation}{https://www.nsf.gov/awardsearch/show-award?AWD_ID=2506333} through award \grantnum{DRL-2506333}{DRL-2506333}.
\end{acks}

\bibliographystyle{ACM-Reference-Format}
\bibliography{main}

\newpage

\appendix
\section{Example Design Argument Template from ARS}
\label{appendix:template}

\begin{figure}[!h]
  \centering
  \includegraphics[width=\linewidth]{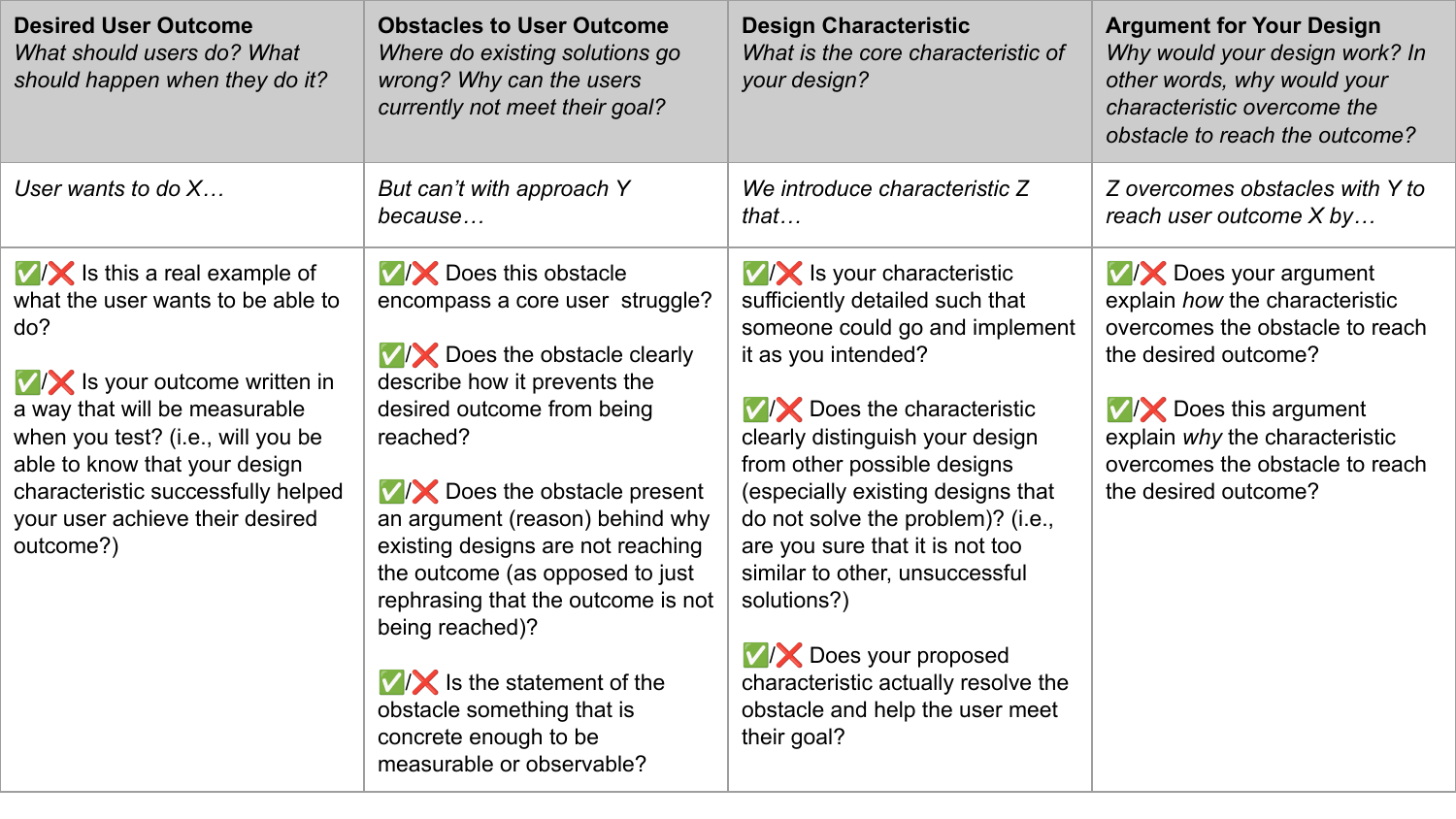}
  \caption{Representations are frequently used in ARS~\cite{zhang2017agile} to support students in their argumentation. Here, we see a Design Argument Template that guides students through user needs, obstacles to meeting those needs, a new design characteristic, and an argument for why the characteristic would be effective. The bottom row includes scaffolding questions to help students assess the quality of their argument.}
  \label{fig:design-argument-template}
  \Description[Four-column design argument worksheet.]{A landscape worksheet organized as a four-column table with three rows, including the header row. The columns are Desired User Outcome, Obstacles to User Outcome, Design Characteristic, and Argument for your design. Each column includes a guiding prompt. The second row includes example sentence stems. The bottom row has self-check questions specific to each column.}
\end{figure}

\section{Reflection Questions by Practice Agent}
\label{appendix:agent-questions}

Students were asked the following questions for each Practice Agent.

\subsection{{[}plan{]}}
No reflection questions were asked for planning-only practices.

\subsection{{[}reflect{]}}

\textit{If the practice was attempted:}
\begin{itemize}
  \item Enter your reflections on the prompt your mentor suggested above.
  \item How did this reflection help you understand how you currently practice and why that happens, and how your practices could change? What was helpful? What obstacles or concerns do you still have?
\end{itemize}

\subsection{{[}self-work{]}}

\textit{If the practice was attempted:}
\begin{itemize}
  \item Share a \textbf{link} to the deliverable that shows what you worked on (e.g., a Google Doc, Figma prototype, GitHub repository, or image). For images, add them to a Google Doc and share the link. Ensure the link is accessible to anyone who has it.
  \item How did your understanding change? What new risk(s) do you see?
  \item What obstacles came up in trying to do it, if any?
\end{itemize}

\textit{If the practice was not attempted:}
\begin{itemize}
  \item What prevented you from doing this work practice? Why did this prevent you from doing it (e.g., lack of time, no longer important after addressing another risk)?
  \item Are there other strategies that you could have tried? For example, could your mentor or your peers have helped you? Why or why not?
\end{itemize}

\subsection{{[}help{]}}

Help-seeking practices varied across contexts.

\subsubsection{Office Hours}

\textit{If the practice was attempted:}
\begin{itemize}
  \item Share a \textbf{link} to an image of what you worked on or discussed at \textbf{Office Hours}. For images, add them to a Google Doc and share the link. Ensure the link is accessible to anyone who has it.
  \item How did \textbf{Office Hours} help progress your understanding? What new risk(s) did it reveal?
  \item What obstacles came up during \textbf{Office Hours}, if any?
\end{itemize}

\textit{If the practice was not attempted:}
\begin{itemize}
  \item Did anything prevent you from attending \textbf{Office Hours} this week? If so, why?
  \item Did anything prevent you from working on the suggested practice at \textbf{Office Hours}? If so, why?
\end{itemize}

\subsubsection{Peer Help}

\textit{If the practice was attempted:}
\begin{itemize}
  \item Share a \textbf{link} to the deliverable that shows what you worked on at \textbf{Peer Help} (e.g., a Google Doc, Figma prototype, GitHub repository, or image). Ensure the link is accessible to anyone who has it.
  \item What did working with a peer help you accomplish? How did that help you progress your sprint?
  \item Were you able to complete your help request? If not, what obstacles came up (e.g., the help request was not sufficiently scoped)?
\end{itemize}

\textit{If the practice was not attempted:}
\begin{itemize}
  \item Did anything prevent you from attending \textbf{Peer Help} this week? If so, why?
  \item Did anything prevent you from working on the suggested practice at \textbf{Peer Help} (e.g., completing an activity to slice a task for peer help)?
\end{itemize}

\subsubsection{Working With Suggested Others}

\textit{If the practice was attempted:}
\begin{itemize}
  \item Share a \textbf{link} to the deliverable that shows what you worked on with the people your mentor suggested (e.g., a Google Doc, Figma prototype, GitHub repository, or image). Ensure the link is accessible to anyone who has it.
  \item What did working with the people your mentor suggested help you accomplish? How did that help you progress your sprint?
  \item Were you able to complete your help request? If not, what obstacles came up (e.g., a new issue emerged during help-seeking)?
\end{itemize}

\textit{If the practice was not attempted:}
\begin{itemize}
  \item Did anything prevent you from asking the people your mentor suggested for help? If so, why?
\end{itemize}

\section{Orchestration Scripts for Computationally Supporting Situated Work Activities}
\label{appendix:os-diagram}
\begin{figure}[h]
  \centering
  \includegraphics[width=\linewidth]{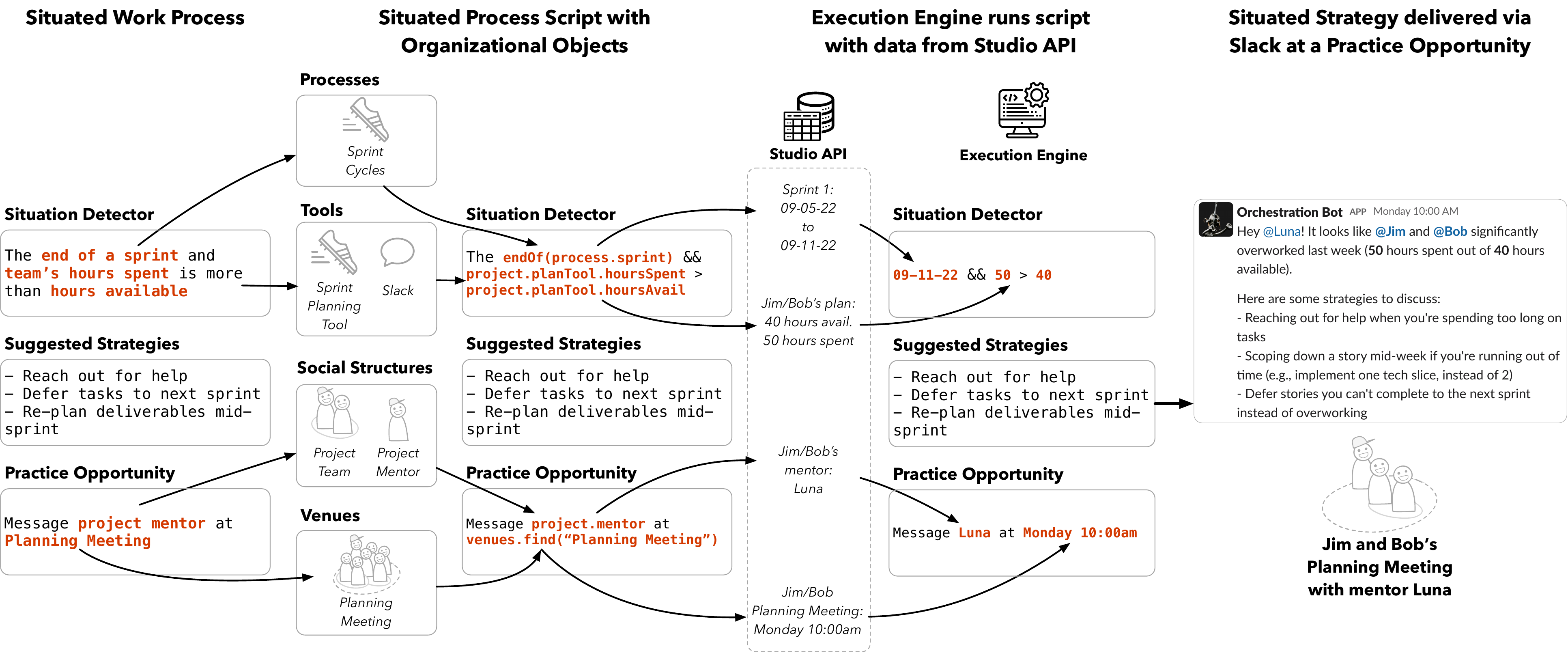}
  \caption{Orchestration Scripts~\cite{garg_orchestraton-scripts_2023} enable computationally encoding Situated Work Processes that organizations want to foster as Situated Process Scripts, such as helping students plan feasible research sprints. These are encoded using Organizational Object abstractions that model an organization's ways of working---their work processes, tools, social structures, and collaboration venues. Scripts encoded using these abstractions are executed by an Execution Engine that pulls data from a Studio API---which contains specific information about the organization---and surfaces strategies at relevant practice opportunities throughout the work process.}
  \label{fig:os-diagram}
  \Description[Architecture flow for Orchestration Scripts.]{A left-to-right architecture diagram with four stages. The first stage represents a Situated Work Process with a situation detector for an infeasible sprint, suggested strategies for reaching out for help or changing scope, and a practice opportunity at a planning meeting. The second stage encodes this as a Situated Process Script with Organizational Objects for sprint cycles (processes); a sprint-planning tool and Slack communication tool (tools); project members and mentors (social structures); and planning meetings (venues). The third stage shows an Execution Engine using Studio API data to resolve specific information for each Organizational Object. The final stage shows a Slack message delivered to the mentor at the practice opportunity, suggesting strategies to discuss with the student.}
\end{figure}

Figure~\ref{fig:os-diagram} describes how Orchestration Scripts~\cite{garg_orchestraton-scripts_2023} enable pre-encoded situated work processes to be computationally facilitated. Our Situated Practice System extends this system to execute on-demand Practice Scripts to support novices' practice development (see Section~\ref{sssec:practice-scripts}).

\section{Codebook for Practice Gaps in Work Issues from Coaches' CAP Notes}
\label{appendix:regulation-gap-coding}

\begin{table*}[ht]
  \caption{Qualitative codes for cognitive regulation gaps. These correspond to instances where the student was lacking the skills to approach problems with an unknown answer or to determine exactly what the problem is.}
  \label{tab:regulation-gaps-cognitive}
  \footnotesize
  \begin{tabular}{@{}p{0.4885\textwidth} p{0.4885\textwidth}@{}}
    \toprule
    \textbf{Regulation Gap Code}                                                                                                                                                                                                                                                                                                      &   
    \textbf{Example for Coders} \\
    \midrule

    \textbf{Representing problem and solution spaces:} The way the student structures or presents the information does not effectively support reasoning, analysis, or communication.                                                                                                                                                 &   
    Jacob struggles to create a representation that would help him show both a working example and one where the system breaks down, so the system's limitations can be understood. \\ \midrule

    \textbf{Assessing risks:} The student struggles with identifying the riskiest risks and/or prioritizing them. They may skip ahead, take on unnecessary or unimportant tasks, or pursue impractical plans because they do not properly assess and prioritize the risks.                                                            &   
    John wasted time jumping ahead by creating multiple very detailed mockups when he should have been focusing on the riskiest risk at hand: identifying his target audience and conducting needfinding. \\ \midrule

    \textbf{Critical thinking and argumentation:} The student struggles to construct well-reasoned arguments supported by evidence or lacks a conceptual understanding of the task at hand. They might find it difficult to identify conceptual differences, treating concepts as too similar when they actually differ meaningfully. &   
    Eloise does not fully understand what a regulation gap is or how to distinguish among the different types, and keeps making the wrong changes to her prototypes as a result. \\

    \bottomrule
  \end{tabular}
  \Description[Two-column codebook table for cognitive regulation gaps.]{A two-column codebook with Regulation Gap Code and Example for Coders. It lists three cognitive regulation gap codes, each with one example.}
\end{table*}

\begin{table*}[ht]
  \caption{Qualitative codes for metacognitive regulation gaps. These correspond to instances in which the student struggled with planning, help-seeking, collaboration, and reflection.}
  \label{tab:regulation-gaps-metacognitive}
  \footnotesize
  \begin{tabular}{@{}p{0.4885\textwidth} p{0.4885\textwidth}@{}}
    \toprule
    \textbf{Regulation Gap Code}                                                                                                                                                                                                                                      &   
    \textbf{Example for Coders} \\
    \midrule

    \textbf{Forming feasible plans:} The student struggles to develop structured, realistic, and actionable plans. This could include what their outcome should be and how to measure their outcome.                                                                  &   
    Penelope overloaded herself with tasks this sprint. While she got a lot of them done, she had limited depth in each because she was spread too thin. \\ \midrule

    \textbf{Planning effective iterations:} The student struggles to create a deliverable that addresses the sprint's riskiest risk. The student may struggle with slicing, breaking larger problems into smaller ones, prioritization, or understanding the problem. &   
    David's prototype included multiple unnecessary features that did not support his research hypothesis, preventing him from testing it this week. \\ \midrule

    \textbf{Leveraging resources and seeking help:} The student struggles to identify and use available materials, expertise from others, and information to enhance their learning and problem-solving.                                                              &   
    Peter was struggling with a programming bug, but decided to keep working on it himself rather than ask his peers for help or resources. \\

    \bottomrule
  \end{tabular}
  \Description[Two-column codebook table for metacognitive regulation gaps.]{A two-column codebook with Regulation Gap Code and Example for Coders. It lists three metacognitive regulation gap codes, each with one example.}
\end{table*}

\begin{table*}[ht]
  \caption{Qualitative codes for emotional regulation gaps. These correspond to instances in which the student had dispositions towards themselves and learning that affect their motivation, cognition, and metacognition.}
  \label{tab:regulation-gaps-emotional}
  \footnotesize
  \begin{tabular}{@{}p{0.4885\textwidth} p{0.4885\textwidth}@{}}
    \toprule
    \textbf{Regulation Gap Code}                                                                                                                                                                   &   
    \textbf{Example for Coders} \\
    \midrule

    \textbf{Fears and anxieties:} The student may have a fear of imperfection, failures, not knowing, etc., which causes them to shy away from the work, and/or not want to try things themselves. &   
    Jennifer had a well-planned sprint to carry out, but got too caught up in designing the solution perfectly rather than building a first prototype. \\ \midrule

    \textbf{Embracing challenges and learning:} The student tries to brute force their way through a solution or runs away from it, rather than thinking about the strategy and approach.          &   
    Riley's system was not producing her optimal output, so she tried to overfit to a single example rather than taking a step back to observe the patterns that caused the system to fail. \\

    \bottomrule
  \end{tabular}
  \Description[Two-column codebook table for emotional regulation gaps.]{A two-column codebook with Regulation Gap Code and Example for Coders. It lists three emotional regulation gap codes, each with one example.}
\end{table*}

\end{document}